\documentclass[aps,prc,reprint,longbibliography,floatfix]{revtex4-1}

\usepackage{graphicx}
\usepackage{hyperref}
\newcommand{\etal}{\textit{et al}.\ }

\begin{document}

\title{Hysteretic vortex matching effects in high-$T_c$ superconductors with nanoscale periodic pinning landscapes fabricated by He ion beam projection technique}

\author{G.~Zechner}

\author{F.~Jausner}

\author{L.~T.~Haag}

\author{W.~Lang}
\email[]{wolfgang.lang@univie.ac.at}

\affiliation{University of Vienna, Faculty of Physics, Electronic Properties of Materials, Boltzmanngasse 5, A-1090, Wien, Austria}

\author{M.~Dosmailov}

\author{M.~A.~Bodea}

\author{J.~D.~Pedarnig}

\affiliation{Johannes-Kepler-University Linz, Institute of Applied Physics, Altenbergerstrasse 69, A-4040 Linz, Austria}

%\date{\today}

\begin{abstract}
Square arrays of sub-micrometer columnar defects in thin YBa$_{2}$Cu$_{3}$O$_{7-\delta}$ (YBCO) films with spacings down to 300~nm have been fabricated by a He ion beam projection technique. Pronounced peaks in the critical current and corresponding minima in the resistance demonstrate the commensurate arrangement of flux quanta with the artificial pinning landscape, despite the strong intrinsic pinning in epitaxial YBCO films. Whereas these vortex matching signatures are exactly at predicted values in field-cooled experiments, they are displaced in zero-field cooled, magnetic-field ramped experiments, conserving the equidistance of the matching peaks and minima. These observations reveal an unconventional critical state in a cuprate superconductor with an artificial, periodic pinning array. The long-term stability of such out-of-equilibrium vortex arrangements paves the way for electronic applications employing fluxons.

\end{abstract}

% insert suggested PACS numbers in braces on next line

\maketitle

\section{Introduction}

Investigations of magnetic vortices in superconductors interacting with periodic pinning arrays have been motivated by the rich diversity of different physical phenomena that result from the interplay of pinning, elastic, and thermal energies and the influences of the dimensionality, anisotropy, and spatial arrangement of vortices  \cite{LYKO93,BAER95,HARA96b,CAST97,LYKO11,CRIS05,OOI05,AVCI10,SILH11,SOCH11,SWIE12,SHAW12,TRAS13,HAAG14, TRAS14,TRAS15,POCC15}. But also, the motion of vortices in a superconductor gives rise to dissipation that is detrimental for most technical applications. Developing efficient strategies to block the mobility of vortices is, thus, one of the high-priority goals of current applied superconductivity research. Tailored artificial pinning defects offer the chance not only to realize enhanced vortex pinning, but also many different ways of flux quanta manipulation, like guided vortex motion \cite{WORD04}, vortex ratchets, \cite{OOI07,PALA12} and valves --- building blocks for fast data manipulation and low-dissipative computing applications \cite{HAST03,MILO07}. An important prerequisite for such undertaking is the realization of stable fluxon arrangements in non-equilibrium positions within the artificial pinning landscape, in analogy to the critical state in unpatterned hard superconductors.

Up to now, primarily metallic superconductors have been used to study regular artificial defects, although cuprate high-$T_c$ superconductors (HTSC) would offer much more convenient cooling requirements. In HTSC, however, the anisotropic layered structure, strong thermal fluctuations and the $d$-wave symmetry of the order parameter introduce additional complexity. Also, the typical average distance of intrinsic defects in  YBa$_{2}$Cu$_{3}$O$_{7-\delta}$ (YBCO) films is about 300~nm \cite{DAM99}. Therefore, the common methods for patterning artificial arrays employed for metallic superconductors, typically allowing for $\mu$m-scale lattices, are only of limited use for HTSC films. Addressing these issues we report on an elegant method for fabricating periodic artificial pinning landscapes into thin YBCO films and on hysteretic commensurability effects between fluxons and these pinning arrays that demonstrate stable non-equilibrium arrangements of flux quanta.

\section{Experimental techniques}

Thin films of YBa$_2$Cu$_3$O$_{7-\delta}$ were grown epitaxially on (100) MgO single-crystal substrates by pulsed-laser deposition using 248~nm KrF-excimer-laser radiation at a fluence of 3.2~J/cm$^2$ resulting in thicknesses of the films of $t_z = (210 \pm 10)$~nm. The critical temperatures of the as-prepared films were $T_c \sim 90$~K, the transition width $\Delta T_c \sim 1$~K, and the critical current density $j_c \sim 3$~MA/cm$^2$ at 77~K and zero magnetic field. For the electrical transport measurements two identical bridges with dimensions $240 \times 60\ \mu \mathrm{m}^2$ were patterned by photolithography and wet chemical etching. Contacts were established in a four-probe geometry using sputtered Au pads with a voltage probe distance of $100\ \mu \mathrm{m}$.

\begin{figure}[t]
\centering
\includegraphics*[width=\columnwidth]{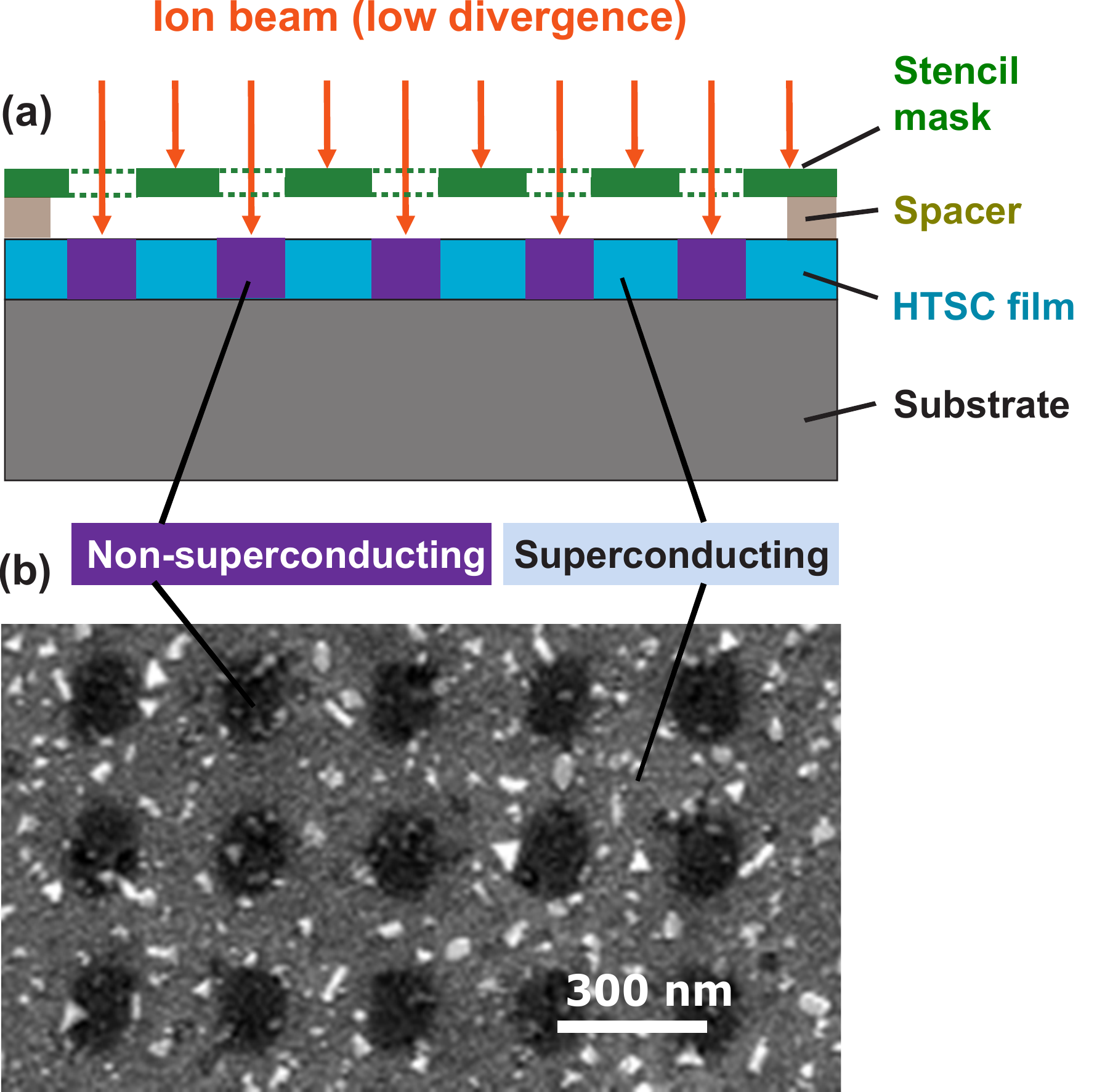}
\caption[]{(a) The principle of masked ion beam direct structuring (MIBS). (b) Scanning electron microscopy (secondary electron detection) picture of the surface of a thin YBCO film after MIBS with 75 keV He$^+$ ion irradiation. The dark regions correspond to an array of defect-rich, non-superconducting nano-cylinders.}
\label{fig:MIBS}       % Give a unique label
\end{figure}

The pinning landscape in the films was created by masked ion beam irradiation (MIBS), sketched in Fig.~\ref{fig:MIBS}(a). It takes advantage of the fact that irradiation of YBCO with He$^+$ ions of moderate energy (75 keV in our experiment) leads to a suppression of the critical temperature $T_c$. This effect is due to a displacement of mainly the chain oxygen atoms, while the skeleton of the crystalline structure remains intact \cite{LANG06a}. A thin Si stencil mask (custom fabricated with e-beam lithography by ims-chips, Germany) was placed on top of the YBCO film and kept at well-defined distance by a circumferential spacer layer of $1.5\ \mu$m-thick photoresist. By this procedure any contact between the surfaces of the mask and the YBCO film was avoided. The parallel alignment of the mask relative to the pre-patterned YBCO bridge was achieved via monitoring in an optical microscope with the help of marker holes in the Si membrane. Three different masks were used. For the main results a square array of about $670 \times 270$\ holes with diameters $D = (180 \pm 5)$~nm and $d = (302 \pm 2)$~nm lattice constant was used, covering the entire YBCO bridge. Additional samples were irradiated through masks with the same hole diameters, but 500~nm and 1~$\mu$m lattice constants, respectively. MIBS provides a 1:1 projection of the hole pattern inscribed in the mask, since the ion beam can only reach the YBCO sample through the mask holes. The ion irradiation produces columnar defect-rich regions (CDs) in the sample \cite{LANG06a}. All other parts of the sample, as well as the electrical contacts, were protected from the irradiation. The second bridge on the substrate was not irradiated and served as a reference.

The irradiations were performed with a fluence of $3 \times 10^{15}\ \textrm{cm}^{-2}$ in a commercial ion implanter (High Voltage Engineering Europa B. V.) with rapid lateral beam scanning and dose monitoring by Faraday cups \cite{LANG06a}. Beam current was kept small and the sample stage was cooled to avoid heating of the YBCO film above room temperature that might have resulted in loss of oxygen. The beam direction was parallel to the sample's $c$-axis direction. Fig.~\ref{fig:MIBS}(b) shows the resulting contrast pattern of the columnar defect array (CDA) at the surface of the YBCO film in a scanning-electron microscopy picture. Note that the dark areas do not represent holes in the sample, but the different escape rate of secondary electrons from the irradiated material \cite{PEDA10}. MIBS leads to some reduction of the critical temperature depending on the width of non-irradiated channels between the CDs, resulting in $T_c \sim 47$~K ($d=302$~nm), $T_c \sim 80$~K ($d=500$~nm), and $T_c \sim 85$~K ($d=1~\mu$m), respectively. A similar behavior was observed in other studies of CDAs that were produced by masked ion irradiation \cite{SWIE12}.

Somewhat different procedures for patterning of HTSC films by ion irradiation have been reported by several groups. They have in common that the mask is directly deposited onto the sample's surface. Either, photoresist is used as the ion-blocking material, with etched holes defined by optical \cite{KAHL98}, e-beam \cite{SWIE12,TRAS13,TRAS14,TRAS15} or focused ion beam lithography \cite{KATZ00}, or a deposited metal layer that is patterned by ion beam milling \cite{KANG02a,BLAM03,BERG05}. In contrast, the present MIBS technology provides many advantages over other methods used for nanopatterning of HTSC: (1) The desired pattern is fabricated in a {\em single-step} process, directly resulting in the as-required modification of specific portions of the material. The extent of this change can be controlled by the ion fluence, leading to superconducting (with reduced $T_c$), normal conducting, or even insulating properties. (2) The method avoids any contact with the sample surface and does not require chemical treatment or etching, thus, preventing possible surface damage. (3) It is a time-economic parallel method applicable to large areas and the mask can be reused many times allowing for scalability in industrial manufacturing. (4) The surface remains essentially flat which permits the preparation of multi-layer structures and avoids deterioration of the film by out-diffusion of oxygen through open side faces.

Magnetoresistance and critical current were measured in a closed-cycle refrigerator with temperature control by a Cernox resistor \cite{HEIN98}. The magnetic field, supplied by an electromagnet and oriented perpendicular to the sample surface, was monitored by a calibrated Hall probe with an accuracy of $\pm 1$~mT. The critical current $I_c(B)$ was measured with two different procedures. To ensure that the vortices can arrange themselves in equilibrium positions throughout the sample, the YBCO film was field-cooled (FC)  from 100~K to the final measurement temperature. After a delay of 5 min to achieve thermal equilibrium an exponential current ramp was executed until the voltage criterion of 100~nV, corresponding to $10 \mu$V/cm, was reached. Subsequently, the temperature was raised again to 100~K, the magnetic field incrementally increased or decreased and the process repeated. Alternatively, the sample was cooled with the magnetic field set to zero within the accuracy $\pm 1$~mT of our gaussmeter (ZFC) and the data collected at the target temperature by ramping the magnetic field without warming up the sample. Corresponding protocols were used to measure the magnetoresistance $R(B)$.

\section{Results}

The commensurate arrangement of vortices within a periodic lattice of pinning regions is commonly demonstrated by ``vortex matching'' effects. For a square array of pins the (first) matching field is
\begin{equation}
\label{eq:matching}
B_m = \frac{{{\phi _0}}}{{{d^2}}},
\end{equation}
where $\phi_0$ is the flux quantum. In principle, vortex matching can also occur at fields $n B_m$ with $n$ any rational number. For convenience we will use $n = 0$ to denote the absence of vortices. The various arrangements of vortices with respect to the defect lattice for integer and fractional values of $n$ have been visualized by Lorentz microscopy in a superconducting Nb film \cite{HARA96b}.

Whereas vortex matching effects in metallic superconductors have been investigated thoroughly with typically $\mu$m-sized antidot structures, \cite{LYKO93,METL94,BAER95,HARA96b,LYKO11,SILH11} few experiments have been reported for HTSC. Peaks of the critical current \cite{CAST97,HAAG14} or minima in the magnetoresistance at $n B_m$ \cite{OOI05,AVCI10,SWIE12,TRAS13,HAAG14} have been observed. Also, vortex ratchet and rectification effects have been reported in YBCO \cite{WORD04,PALA12}, as well as geometric frustration effects in non-uniform pinning arrays \cite{TRAS14,TRAS15}.

\begin{figure}[t]
\centering
\includegraphics*[width=\columnwidth]{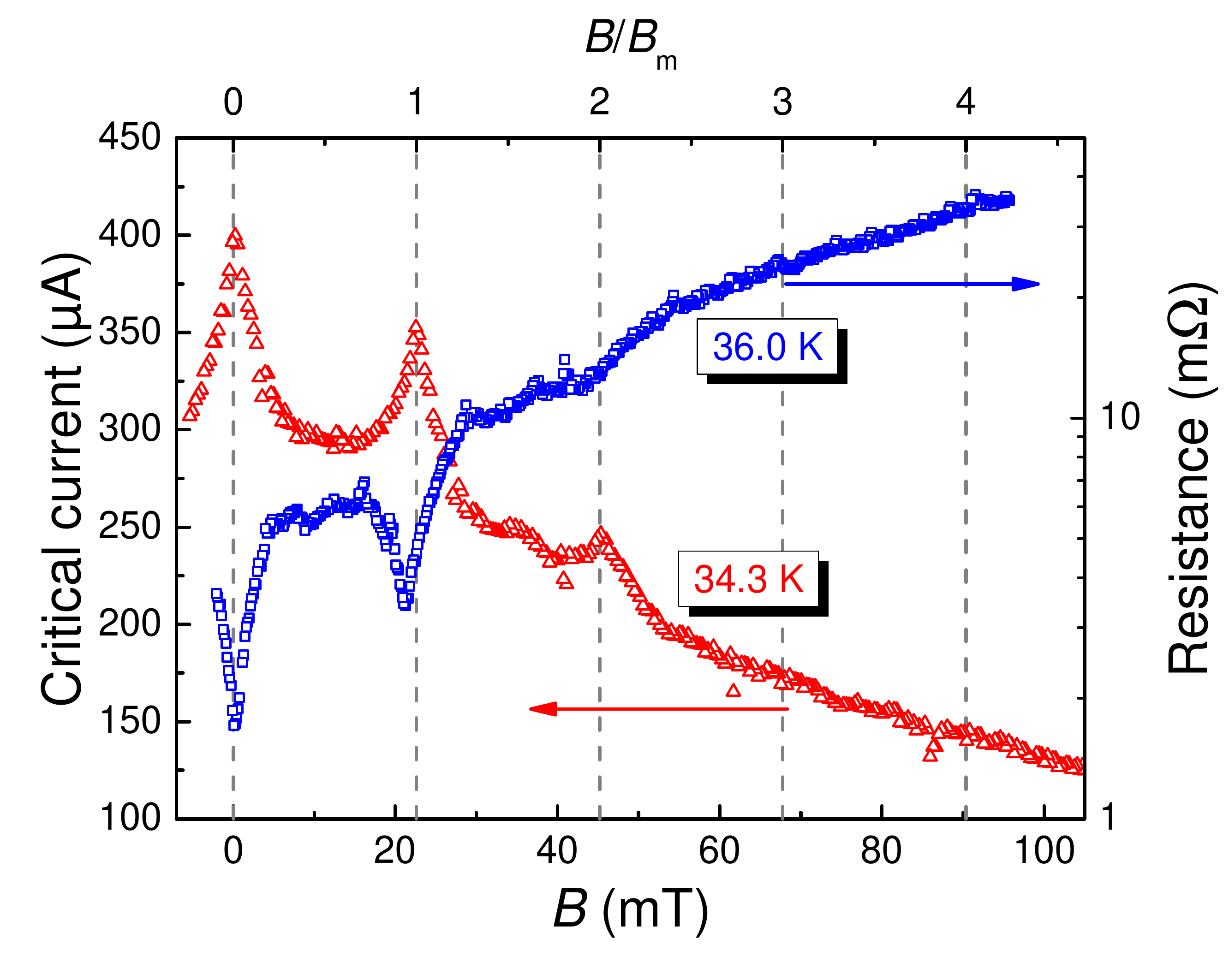}
\caption[]{Critical current and magnetoresistance as a function of the applied magnetic field $B$ after field cooling the sample from $T=100\ \mathrm{K}$ for every data point. The upper horizontal axis is scaled to multiples of the matching field $B_m = 22.6$~mT.}
\label{fig:JcMR}       % Give a unique label
\end{figure}

In Fig.~\ref{fig:JcMR} two different experimental tests for vortex matching effects in the FC case are demonstrated. The critical current $I_c(B)$ at a temperature $T = 34.3$~K, corresponding to a reduced temperature $t = T/T_c = 0.73$, probes the static situation, where vortices are pinned, and the magnetoresistance $R(B)$ at $T = 36.0$~K ($t = 0.77$) the dynamic case when vortices move. The distinct maxima in $I_c(B)$ and the minima of $R(B)$ are positioned exactly at multiples of the matching field according to Eq.~(\ref{eq:matching}), where a commensurate relation exists between the vortex lattice and the CDA defined by the geometrical parameters of the stencil mask. Calculations in the framework of nonlinear Ginzburg-Landau theory predict a similar behavior of $I_c(B)$ for equilibrium vortex arrangements \cite{BERD06}. Careful inspection reveals signatures of vortex matching at $n = \frac{1}{2}$ and $n = \frac{3}{2}$, too. Naturally, with the absence of an external magnetic field, $I_c$ attains a maximum and $R$ a minimum.

The characteristic peaks and minima become less significant, if the lattice spacing of the CDA is increased. In the $d = 500\ $nm sample, they are still well visible, whereas they are almost washed-out in the $d = 1\ \mu$m sample. This observation is in line with previous reports, where weak cusps in $I_c(B)$ have been identified in a YBCO film perforated with a square array of holes \cite{CAST97} with $d = 1\ \mu$m and in a Bi$_2$Sr$_2$CaCu$_2$O$_8$ (BSCCO) ribbon with $d = 0.5\ \mu$m \cite{AVCI10}. Conversely, in a YBCO film with $d = 120$~nm a significant peak of $I_c(B)$ at $B_m$ has been seen \cite{SWIE12}. We attribute the pronounced matching effects in our sample to the fact that the lattice constant of the CDA is comparable to the typical intrinsic defect spacing in epitaxial YBCO films.

\begin{figure}[t]
\centering
\includegraphics*[width=\columnwidth]{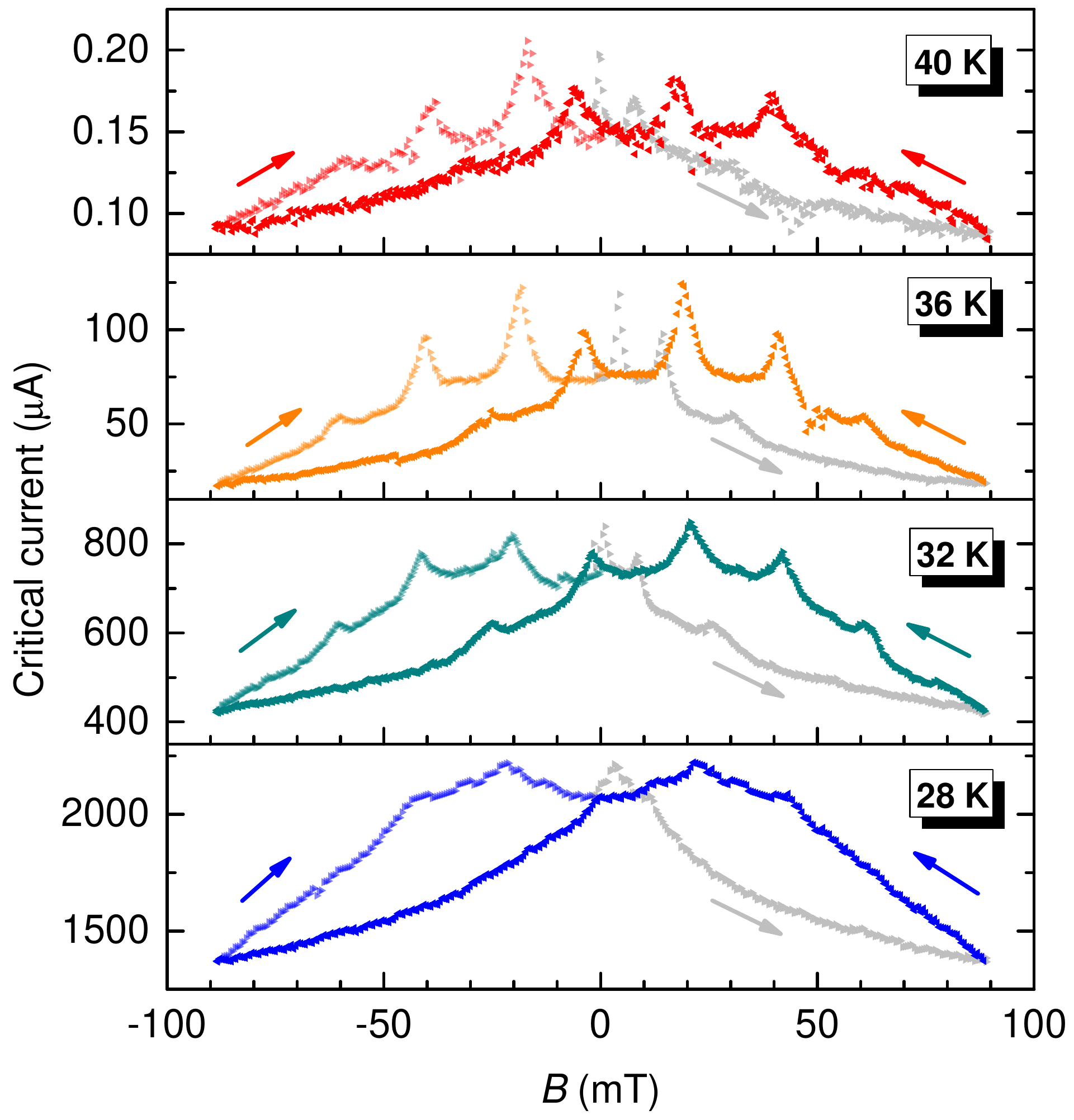}
\caption[]{Hysteretic behavior of the critical current after zero-field cooling at the respective temperature and ramping the applied magnetic field $B$ through a full cycle. Grey symbols denote the virgin curve.}
\label{fig:Tvar}       % Give a unique label
\end{figure}

\begin{figure}[t]
\centering
\includegraphics*[width=\columnwidth]{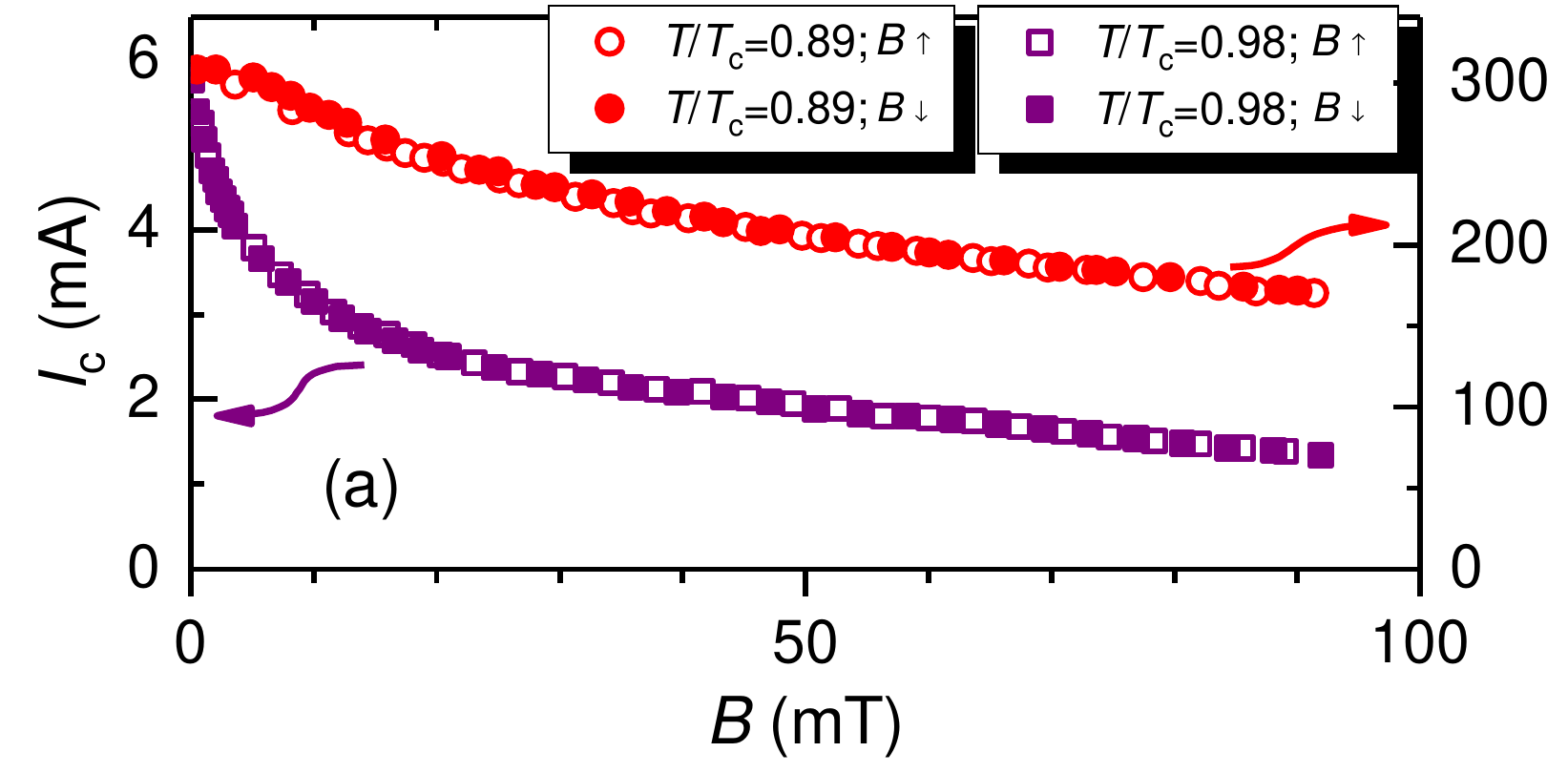}
\includegraphics*[width=\columnwidth]{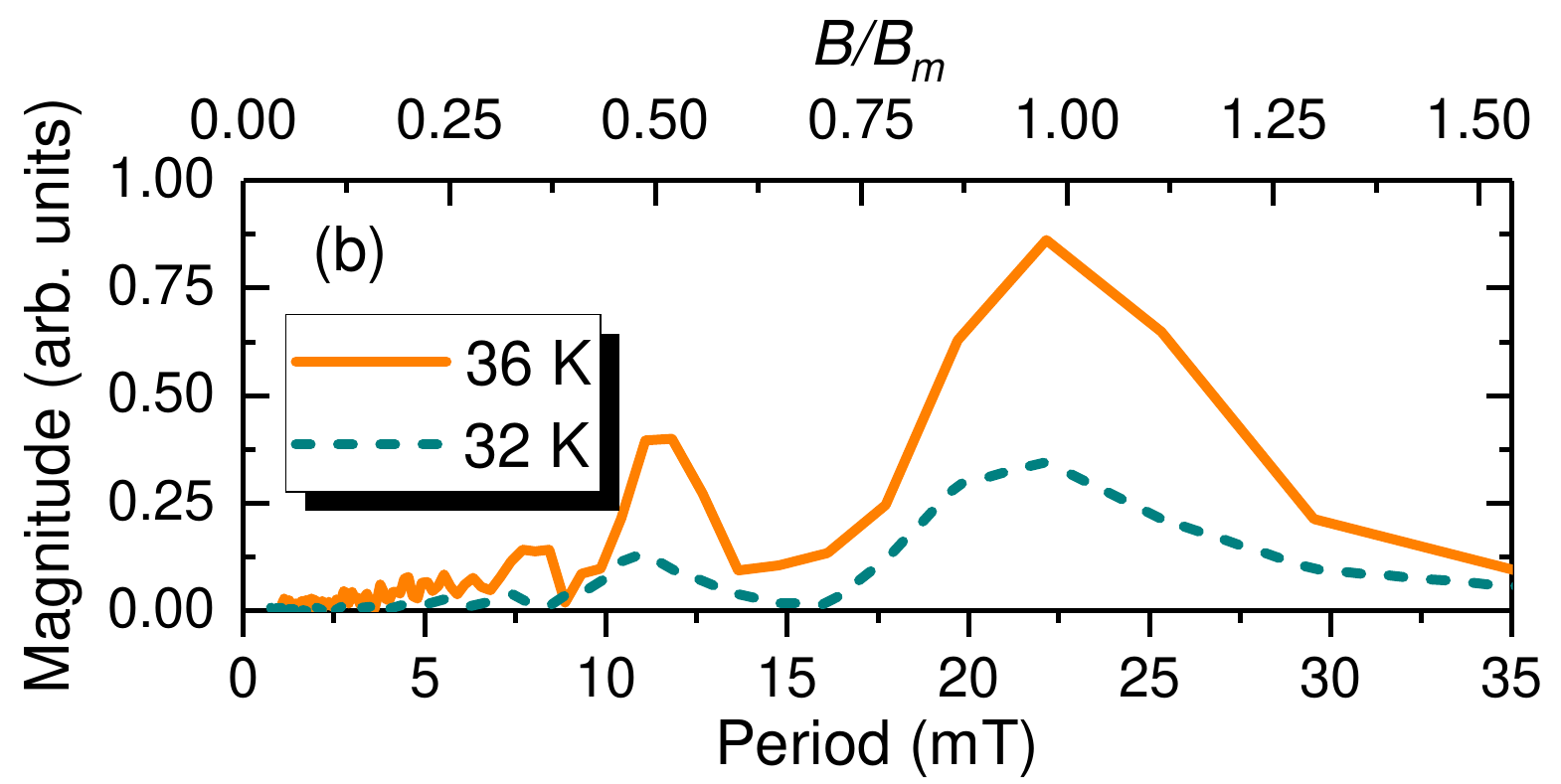}
\caption[]{(a) Critical current of the unirradiated reference bridge after ZFC in increasing (open symbols) and decreasing (full symbols) magnetic field at $t = 0.89$ (circles) and $t = 0.98$ (squares), respectively. (b) Periodic components in the down-ramped branches from Fig.~\ref{fig:Tvar} determined via Fourier transform.}
\label{fig:FFT}       % Give a unique label
\end{figure}

A strikingly different picture arises in the ZFC experiments. Fig.~\ref{fig:Tvar} presents $I_c(B)$ of the patterned YBCO film for various temperatures after cooling from $T = 100$~K and then cycling the magnetic field. The grey branches represent the virgin curves, the bright colored ones the decreasing-field ramp, and the light-colored branch returns to $B=0$. We emphasize that a complementary procedure, in which the sample was first cooled below $T_c$ in a field of 90~mT and then the magnetic field reduced, leads to the same curve  as the ramped-down branch of the ZFC experiment in Fig.~\ref{fig:Tvar}. All curves display a strong hysteresis during these cycles, in sharp contrast to most of the other related experiments that did not report hysteretic effects of electrical transport properties \cite{OOI05,AVCI10,SOCH11,SWIE12,TRAS13,TRAS14,TRAS15,POCC15}. Only a tiny hysteresis of the critical current was reported in perforated Sn \cite{LYKO93,LYKO11} and YBCO films \cite{CAST97} with $1\ \mu$m inter-hole distance. Fig.~\ref{fig:FFT}(a) demonstrates that irreversible effects are also absent in our unirradiated reference bridge.

Whereas in the virgin curves the peaks seem to appear at irregular positions that cannot be correlated to matching fields, the decreasing-field data look very similar to the FC data in Fig.~\ref{fig:JcMR}, but shifted by a magnetic field $B_{shift}=(19 \pm 2)$~mT. Thus, the field-ramped curves reveal the highest $I_c$ at non-zero field. In addition, the distance between the $n$ and $n+1$ peaks in $I_c$ corresponds to $B_m$ at all temperatures, despite a large variation of critical current and hysteresis area of the curves. Remarkably, even $B_{shift}$ varies only slightly with temperature and the peaks can be observed over a wide temperature range $0.60 < t < 0.85$.

\begin{figure}[t]
\centering
\includegraphics*[width=\columnwidth]{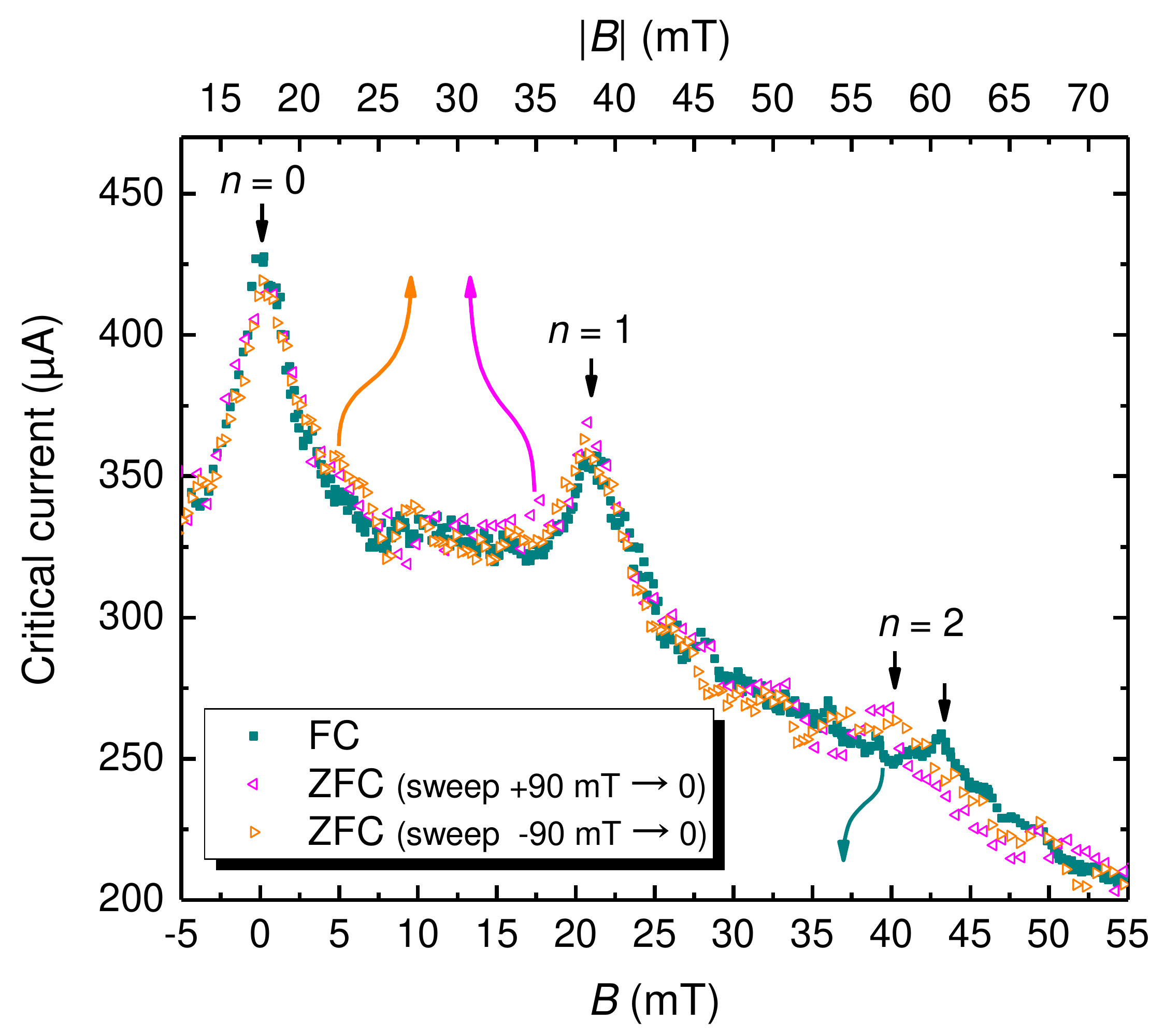}
\caption[]{Comparison of the shape of the matching peaks in zero-field cooled (ZFC) and field cooled (FC) experiments at 35~K. The ZFC data are taken from the two branches of the hysteresis loop when the magnetic field was reduced. The displacement of the upper scale (ZFC data) reflects the shift of the peaks due to the irreversible behavior.}
\label{fig:Peaks}       % Give a unique label
\end{figure}

This periodicity is better demonstrated by Fourier transforms of the down-ramped curves that reveal pronounced maxima at $B/B_m = n = 1$ and smaller peaks at $n = \frac{1}{2}$, see Fig.~\ref{fig:FFT}(b). A direct comparison of the FC and ZFC data collected in subsequent runs at the same temperature is presented in Fig.~\ref{fig:Peaks}. As might be anticipated, the branches swept down from both polarities of the magnetic field are nearly identical. But intriguingly, also the FC data match almost perfectly, aside from a small offset of the $n=2$ peak. These almost identical peak values and shapes indicate that the $I_c$ peaks in both FC and ZFC experiments must be caused by similar domains in which vortices are commensurate with the CDA.

\begin{figure}[t]
\centering
\includegraphics*[width=\columnwidth]{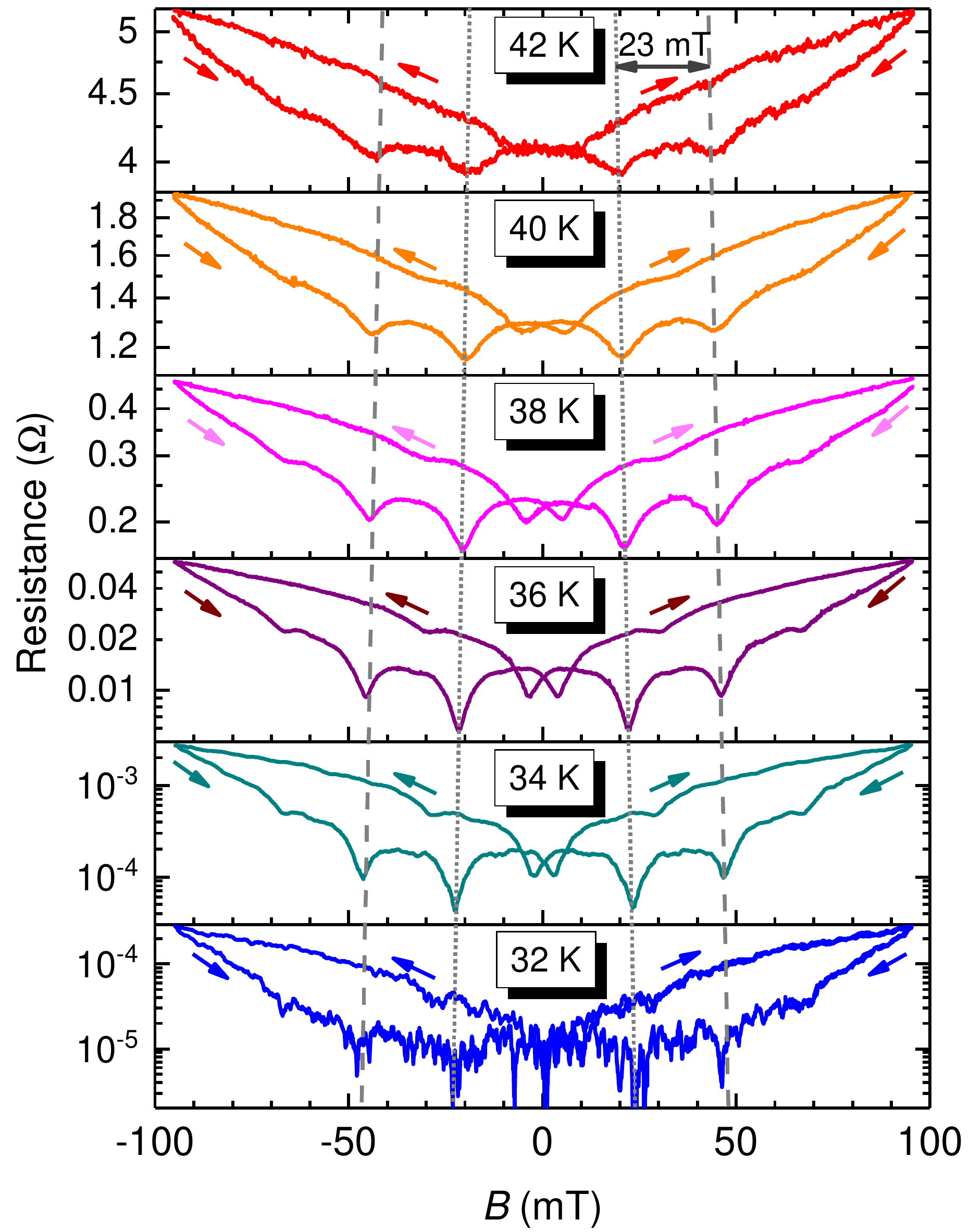}
\caption[]{Hysteretic behavior of the resistance after zero-field cooling at the respective temperature and ramping the applied magnetic field $B$ through full cycles. The second cycle is shown to demonstrate the symmetry of the hysteresis and to exclude the virgin curve. The dotted lines represent the resistance minima corresponding to the $B=0$ state and the dashed lines those attributed to the first matching field, respectively.}
\label{fig:TvarR}       % Give a unique label
\end{figure}

An equivalent hysteresis with the same displacement $B_{shift}$ of the matching peaks can be observed in the magnetoresistance measurements presented in Fig.~\ref{fig:TvarR}. The temperature region where the hysteretic effects can be clearly observed extends close to $T_c$ and into a regime of finite resistance and partial depinning of vortices. Another finding resulting from comparing the ZFC $I_c(B)$ and the $R(B)$ measurements is that $B_{shift}$ is independent of the time-scale of the measurements. Collecting the magnetoresistance data for a full cycle takes about 24 hours, whereas the sweep for the $I_c$ measurement takes 2 to 4 days. Thus, the hysteretic effect is very robust over long time scales.

\section{Discussion}

Beforehand, it is useful to recall the particular geometry of our thin-film samples and relate them to superconducting parameters. The sample's thickness $t_z$ is of the order of the zero-temperature in-plane London penetration depth $\lambda_{ab}(0) \sim 220$~nm in YBCO \cite{LEE93}, length and width are $l > w \gg \lambda_{ab}$. Hence, only marginal curvature of the vortices and their alignment parallel to the crystallographic $c$ axis is expected for $B \parallel c$ in our experiments. However, strong demagnetization effects will be present and the critical state will deviate significantly from a simple Bean model picture \cite{BRAN93c}. The diameter $D$ of the defect columns and their lattice constant are also comparable to $\lambda_{ab}(0)$. The zero-temperature in-plane Ginzburg-Landau coherence length $\xi_{ab}(0) \sim 1.4~\mathrm{nm} \ll D$ \cite{LANG95e}, thus, allowing the accommodation of multiple fluxons in a DC \cite{BUZD93}.

An irreversible behavior of $I_c(B)$ (without vortex matching effects) has been observed in granular YBCO and interpreted by a critical state together with trapped flux in the intergrain voids \cite{EVET88}. The related hysteretic behavior of the microwave surface resistance has been explained by a two-level critical state model, presuming a local flux-density gradient within the grains that is different from the macroscopic one of the entire sample \cite{JI93}. Remarkably strong flux gradients have been reported in BSCCO crystals within an area that was patterned with a blind hole array, but not in the neighbouring pristine parts of the crystal \cite{SHAW12}. Recently, a hysteretic behavior of the Josephson supercurrent in coplanar Al-graphene-Al junctions upon cycling the magnetic field has been reported \cite{MASS16}.

Geometric effects on the critical state of thin homogeneous superconducting strips, similar to the shape of our samples, have been evaluated by Brandt and Indenbom \cite{BRAN93c}. They have calculated the current density and the magnetic field when both a perpendicular magnetic field and a transport current are applied together. The current density reaches its critical value $j_c$ at one edge of the sample, falling off rapidly towards the center of the sample but remaining finite in the entire superconductor. The magnetic field exhibits a curved profile with a larger gradient near the sample's edge. Although these findings seem to complicate further considerations, they nevertheless prove that the critical state in our thin samples can be qualitative interpreted within the common critical state models.

A possible scenario for the unconventional hysteretic shift of the matching peaks in our samples could be the trapping of a remanent field within the superconducting YBCO film due to the large intrinsic pinning of interstitial vortices that adds to the effects of vortex matching in the CDA. In fact, local scanning Hall probe measurements in perforated YBCO films revealed a hysteresis of magnetization between the holes \cite{CRIS05}. However, a comparison of our samples with different $d$ demonstrates that wider lattices lead to a reduction of the hysteresis area and of $B_{shift}$, from $B_{shift} \sim 19$~mT ($d = 302$~nm) to 2.5~mT ($d = 500$~nm) and 0.2~mT ($d = 1\ \mu$m) at $t = 0.98$. In the unpatterned reference YBCO bridge no hysteresis is observed, as demonstrated in Fig.~\ref{fig:FFT}(a). Hence, the appearance of hysteresis appears to be connected with a sufficiently dense CDA and not with pinning of interstitial vortices in the superconducting material in between the CDs.

In fact, in our system two different vortex pinning mechanisms are operative. On the one hand, fluxons are anchored by the normal cores of the CDs and this pinning potential prevents them for entering the superconducting material in between the CDs. Eventually, they can hop from one CD to a neighboring one \cite{SORE17}. On the other hand, interstitial vortices are pinned by intrinsic defects in the unirradiated parts of the sample. Then, the peaks in $I_c$ arise from the fact that a larger number of fluxons is trapped in the CDs, and, conversely, the number of interstitial vortices is smaller. In fact, numerical simulations by Reichardt \etal \cite{REIC98} have predicted a complex dynamical behavior of moving vortices, including their one-dimensional channeling between periodic pinning arrays. The interplay between these two different pinning mechanisms has been also adressed by Avci \etal \cite{AVCI10} for perforated BSCCO films and by Trastoy \etal \cite{TRAS13} for ion-beam patterned YBCO. However, both studies did not report hysteretic transport properties.

To understand the irreversible behavior observed in our ZFC experiments, we propose that the magnetic field gradient inside our sample is mainly established by the strong trapping of fluxons in the CDA, whereas the onset of channelling of interstitial vortices limits the critical current to its rather low value. Thus, unlike in the Bean critical state model for homogeneous superconductors, increasing the current up to $I_c$ does not propagate the magnetic flux towards the sample's center. A much higher Lorentz force would be needed to unpin the fluxons locked at the CDA. The critical current is determined only by the interstitial vortices in a small zone at the sample's edge and is independent of the flux profile within the sample. In fact, Fig.~\ref{fig:Peaks} reveals that the size and shape of the peaks are almost identical in FC and ZFC measurements and, thus, confirms that $I_c$ must be dominated by similar mechanisms despite of different flux profiles.

Now, the strikingly robust distance between the matching peaks/minima in ZFC measurements needs to be addressed. We presume that the critical state splits into uniform domains, in which every CD is populated by the same number $n$ of fluxons, and neighbouring regions by $n \pm 1$. Indeed, a terraced critical state was proposed for vortex-pinning lattices by Cooley and Grishin \cite{COOL95}, where in circumferential regions of the sample the pinning centers are occupied by multi-vortices, with the nucleation of new terraces upon raising the magnetic field. However, their model is based on a thin superconducting slab in parallel magnetic field without demagnetization effects. Although demagnetization would change the profile of magnetic induction to a more singular behavior near the edges of our samples, this effect is softened by a `smearing' to the order of the film thickness $t_z$ \cite{BRAN93c}. Thus, a quantitative comparison of the terrace model's predictions with our experiments is not possible but intriguing parallels on a qualitative level are obvious and will be discussed below.

A key prediction of the terrace model is the periodicity of physical properties with the matching field when the magnetic field is ramped. Such were reported for the pinning force in NbTi composite wires \cite{COOL94} and the magnetization of Pb/Ge multilayers with a square lattice of submicron holes \cite{BAER95}. Another important feature of the terraced critical state is that the sample's magnetization is uniform within such a domain and has a strong gradient between neighboring terraces. The current in the sample stratifies into streamlines along the terrace edges, but vanishes within a terrace \cite{COOL95}.

For an estimate of the situation in our sample, we recall that matching effects are best seen in the down-ramps of ZFC cycled fields and the additional observation that $I_c(B)$ curves are the same when the sample is initially FC at 90~mT and then data taken during down ramp. Thus, we can conclude that at 90~mT the sample is field saturated in both experiments. Upon down sweep, a field gradient builds up, eventually leading to $B = B_m$ at the sample's edges and $B \lesssim 90$~mT $\sim 4 B_m$ in its center. If we assume a terraced critical state with integer fillings of $n B_m$ it follows that about four terraces $n = 1 \ldots 4$ can be present at the $n = 1$ peaks in Fig.~\ref{fig:Tvar}. The width of a terrace roughly scales with $n$ and is connected to $\lambda_{ab}$ \cite{COOL95}. The width of the $n = 1$ terrace, thus, should be of the order of $w_{t1} \sim (w/2)/\sum_{n=1}^4 n = 3~\mu$m $\sim 10 \ d$ and be slightly temperature dependent. However, due to demagnetization effects, the terrace located at the sample's edge is probably somewhat smaller.

In an ideal terraced critical state every current streamline should be confined to a single unirradiated channel between two adjacent rows of CDs at the terrace edge with width $w_s = d - D \sim 120$~nm. Since the vorticity between such neighboring CDs changes by $\pm \phi_0$, the local current density in a channel would be $j_s = \phi_0/(\mu_0 d^2 w_s) \sim 15$~MA/cm$^2$. Our YBCO films have a $j_c (t = 0.77) \sim 7.2$~MA/cm$^2 \sim j_s / 2$ and $j_c (t = 0.85) \sim 3.8$~MA/cm$^2 \sim j_s / 4$, which indicates that the width of a current carrying streamline will extend to at least $2d$ ($4d$) for the 36~K (40~K) data in Fig.~\ref{fig:Tvar}. Within such streamlines that encompass several rows of CDs an incommensurate vortex placement is expected.

Our estimates can be compared to a two-dimensional dynamic simulation of vortex arrangements in a critical state with periodic arrays of pinning sites by Reichardt \etal \cite{REIC97}. They found that upon increasing the magnetic field from zero, a commensurate vortex lattice appears at the edge of the sample that persists for a finite range of the external magnetic field around the first matching peak. Upon further increasing $B$ the commensurate terrace is pushed inside the sample and the edge region disorders before forming the next commensurable plateau. Remarkably, at higher matching fields, no terraces were found, but rather stripes and islands of constant $B$, surrounded by current-carrying strings. Also, scanning Hall probe measurements in a perforated Pb thin film indicate a terraced critical state with increasing disorder at higher fields \cite{SILH11}.

We propose a similar situation in our experiments. Since only matching peaks with $n \leq 2$ are visible, we conclude that the critical state in our sample features commensurate stripes near the sample's edge that are separated by more disordered current-carrying streamlines and that the terraced critical state deteriorates at higher $n$. In the framework of such model the peaks in $I_c$ appear when a fully developed terrace occupies the edge of the sample and, hence, only few interstitial vortices are available to be moved by the applied current. Upon increasing or decreasing the field, a disordered state with higher density of interstitial vortices appears, reducing $I_c$, until the next terrace is formed. In addition, demagnetization effects are less important, allowing for wider commensurate domains at the edge, when the field is ramped down and the external magnetic field can partially penetrate the interior of the sample.

The observed matching effects are remarkably robust over the experimentally accessible temperature range. The hysteretic shift of the peaks/minima $B_{shift}$ changes only little with $T$, $B_{shift}(32~\mathrm{K})/B_{shift}(42~\mathrm{K}) \sim 1.2$, less than   $\lambda_{ab}(42~\mathrm{K})/\lambda_{ab}(32~\mathrm{K}) \sim 1.5$, and is therefore considered a nonrelevant effect. Obviously, the matching effects disappear at high temperatures, when pinning at the CDA becomes negligible. More interesting is the behavior at low temperatures. Here, two effects can be considered. First, the intrinsic pinning becomes strong and interstitial vortices are immobilized, straightening out the matching peaks. Second, the self-field of $I_c$ at 28~K reaches $B_{sf}^{max} = \mu_0 I_c / t_z \sim 13\ \mathrm{mT} \sim B_m/2$ and can displace or erode the terraces.

The strict periodicity of the $I_c$ peaks in the FC experiment (Fig.~\ref{fig:JcMR}) is evident from the matching condition of Eq.~(\ref{eq:matching}) but not so straightforward to understand in the ZFC magnetic-field cycled experiments (Figs.~\ref{fig:Tvar}, \ref{fig:TvarR}), where it was observed over a wide temperature range. In fact, the terraced critical state model \cite{COOL95} predicts the formation of a new terrace with $n+1$ fluxon occupation per CD at the sample's edge, when the field is increased by $\Delta B_{n, n+1} = B_m$ and, vice versa a corresponding depletion of terraces when the field is reduced. However, it follows from Eq.~(\ref{eq:matching}) that one flux quantum per CD must be added or removed over the \emph{entire} sample area. The virgin curves in Fig.~\ref{fig:Tvar} demonstrate that this is indeed not the case when magnetic flux enters into the sample from the edge after ZFC and the CDs in the inner parts of the sample are not yet populated by fluxons, thus $\Delta B_{n, n+1} < B_m$ ($n = 0$ or 1). Conversely, after field-saturating the sample and then ramping down, a monotonically decreasing critical flux profile establishes from the center to the edge of the sample, so that by extracting fluxons from the outermost terrace, a redistribution (and extraction) in the entire sample takes place. This scenario also holds when terraces with higher $n$ inside the sample are partially disordered or do not exist at all. A similar periodicity and temperature independence has been also observed in magnetization measurements in perforated Pb/Ge multilayers \cite{BAER95}.

The observed hystereses in $I_c$ (Fig.~\ref{fig:Tvar}) and $R$ (Fig.~\ref{fig:TvarR}) have their origin in the different magnetization profiles during up and down ramps, respectively, which result in different occupation numbers of the outermost terrace at the sample's edge at the same external magnetic field. Since the current will concentrate at the sample's edge, the outermost terrace determines the transport properties and gives rise to matching peaks shifted away from their equilibrium positions.

To summarize, a terraced critical state or a similar commensurate vortex domain structure at the sample's edge provides a plausible scenario for our observations, in particular for the equidistance of the matching peaks, when $I_c$ is determined by the vortex arrangement at the sample's fringes. Evidently, for in-depth understanding of these novel effects, additional theoretical efforts and simulations are required.

\section{Conclusions}

We have demonstrated the fabrication of sub-$\mu$m CDAs in thin YBCO films with a novel, non-contact, single-step ion irradiation technique. The method features a 1:1 reproduction of a periodic structure from a perforated Si stencil mask and has the potential to be further developed into an ion-projection technique with ion-optical reduction \cite{EDER12}. Strong commensurability effects were observed between flux quanta and the CDA in both equilibrium and non-equilibrium fluxon arrangements. They manifest themselves in pronounced maxima of the critical current and minima of the magnetoresistance. In ZFC and magnetic-field ramped data the matching pattern is displaced, but the equidistance between matching peaks conserved, what could be an indication of an unconventional --- probably terraced --- critical state that calls for further investigations. Finally, our results demonstrate an ordered out-of-equilibrium arrangement of flux quanta with long-term stability in a superconductor and pave the way for envisaged fluxonic devices.

\begin{acknowledgments}

We appreciate the help of Klaus Haselgr\"ubler with the ion implanter. MD acknowledges support from the Erasmus Mundus program. The work was supported by the the COST Action MP-1201.

\end{acknowledgments}


\begin{thebibliography}{46}%
\makeatletter
\providecommand \@ifxundefined [1]{%
 \@ifx{#1\undefined}
}%
\providecommand \@ifnum [1]{%
 \ifnum #1\expandafter \@firstoftwo
 \else \expandafter \@secondoftwo
 \fi
}%
\providecommand \@ifx [1]{%
 \ifx #1\expandafter \@firstoftwo
 \else \expandafter \@secondoftwo
 \fi
}%
\providecommand \natexlab [1]{#1}%
\providecommand \enquote  [1]{``#1''}%
\providecommand \bibnamefont  [1]{#1}%
\providecommand \bibfnamefont [1]{#1}%
\providecommand \citenamefont [1]{#1}%
\providecommand \href@noop [0]{\@secondoftwo}%
\providecommand \href [0]{\begingroup \@sanitize@url \@href}%
\providecommand \@href[1]{\@@startlink{#1}\@@href}%
\providecommand \@@href[1]{\endgroup#1\@@endlink}%
\providecommand \@sanitize@url [0]{\catcode `\\12\catcode `\$12\catcode
  `\&12\catcode `\#12\catcode `\^12\catcode `\_12\catcode `\%12\relax}%
\providecommand \@@startlink[1]{}%
\providecommand \@@endlink[0]{}%
\providecommand \url  [0]{\begingroup\@sanitize@url \@url }%
\providecommand \@url [1]{\endgroup\@href {#1}{\urlprefix }}%
\providecommand \urlprefix  [0]{URL }%
\providecommand \Eprint [0]{\href }%
\providecommand \doibase [0]{http://dx.doi.org/}%
\providecommand \selectlanguage [0]{\@gobble}%
\providecommand \bibinfo  [0]{\@secondoftwo}%
\providecommand \bibfield  [0]{\@secondoftwo}%
\providecommand \translation [1]{[#1]}%
\providecommand \BibitemOpen [0]{}%
\providecommand \bibitemStop [0]{}%
\providecommand \bibitemNoStop [0]{.\EOS\space}%
\providecommand \EOS [0]{\spacefactor3000\relax}%
\providecommand \BibitemShut  [1]{\csname bibitem#1\endcsname}%
\let\auto@bib@innerbib\@empty
%</preamble>
\bibitem [{\citenamefont {Lykov}(1993)}]{LYKO93}%
  \BibitemOpen
  \bibfield  {author} {\bibinfo {author} {\bibfnamefont {A.~N.}\ \bibnamefont
  {Lykov}},\ }\bibfield  {title} {\enquote {\bibinfo {title} {Pinning in
  superconducting films with triangular lattice of holes},}\ }\href {\doibase
  10.1016/0038-1098(93)90103-t} {\bibfield  {journal} {\bibinfo  {journal}
  {Solid State Commun.}\ }\textbf {\bibinfo {volume} {86}},\ \bibinfo {pages}
  {531} (\bibinfo {year} {1993})}\BibitemShut {NoStop}%
\bibitem [{\citenamefont {Baert}\ \emph {et~al.}(1995)\citenamefont {Baert},
  \citenamefont {Metlushko}, \citenamefont {Jonckheere}, \citenamefont
  {Moshchalkov},\ and\ \citenamefont {Bruynseraede}}]{BAER95}%
  \BibitemOpen
  \bibfield  {author} {\bibinfo {author} {\bibfnamefont {M.}~\bibnamefont
  {Baert}}, \bibinfo {author} {\bibfnamefont {V.~V.}\ \bibnamefont
  {Metlushko}}, \bibinfo {author} {\bibfnamefont {R.}~\bibnamefont
  {Jonckheere}}, \bibinfo {author} {\bibfnamefont {V.~V.}\ \bibnamefont
  {Moshchalkov}}, \ and\ \bibinfo {author} {\bibfnamefont {Y.}~\bibnamefont
  {Bruynseraede}},\ }\bibfield  {title} {\enquote {\bibinfo {title} {Composite
  flux-line lattices stabilized in superconducting films by a regular array of
  artificial defects},}\ }\href {\doibase 10.1103/physrevlett.74.3269}
  {\bibfield  {journal} {\bibinfo  {journal} {Phys. Rev. Lett.}\ }\textbf
  {\bibinfo {volume} {74}},\ \bibinfo {pages} {3269} (\bibinfo {year}
  {1995})}\BibitemShut {NoStop}%
\bibitem [{\citenamefont {Harada}\ \emph {et~al.}(1996)\citenamefont {Harada},
  \citenamefont {Kamimura}, \citenamefont {Kasai}, \citenamefont {Matsuda},
  \citenamefont {Tonomura},\ and\ \citenamefont {Moshchalkov}}]{HARA96b}%
  \BibitemOpen
  \bibfield  {author} {\bibinfo {author} {\bibfnamefont {K.}~\bibnamefont
  {Harada}}, \bibinfo {author} {\bibfnamefont {O.}~\bibnamefont {Kamimura}},
  \bibinfo {author} {\bibfnamefont {H.}~\bibnamefont {Kasai}}, \bibinfo
  {author} {\bibfnamefont {T.}~\bibnamefont {Matsuda}}, \bibinfo {author}
  {\bibfnamefont {A.}~\bibnamefont {Tonomura}}, \ and\ \bibinfo {author}
  {\bibfnamefont {V.~V.}\ \bibnamefont {Moshchalkov}},\ }\bibfield  {title}
  {\enquote {\bibinfo {title} {Direct observation of vortex dynamics in
  superconducting films with regular arrays of defects},}\ }\href {\doibase
  10.1126/science.274.5290.1167} {\bibfield  {journal} {\bibinfo  {journal}
  {Science}\ }\textbf {\bibinfo {volume} {274}},\ \bibinfo {pages} {1167}
  (\bibinfo {year} {1996})}\BibitemShut {NoStop}%
\bibitem [{\citenamefont {Castellanos}\ \emph {et~al.}(1997)\citenamefont
  {Castellanos}, \citenamefont {W\"ordenweber}, \citenamefont {Ockenfuss},
  \citenamefont {v.d. Hart},\ and\ \citenamefont {Keck}}]{CAST97}%
  \BibitemOpen
  \bibfield  {author} {\bibinfo {author} {\bibfnamefont {A.}~\bibnamefont
  {Castellanos}}, \bibinfo {author} {\bibfnamefont {R.}~\bibnamefont
  {W\"ordenweber}}, \bibinfo {author} {\bibfnamefont {G.}~\bibnamefont
  {Ockenfuss}}, \bibinfo {author} {\bibfnamefont {A.}~\bibnamefont {v.d.
  Hart}}, \ and\ \bibinfo {author} {\bibfnamefont {K.}~\bibnamefont {Keck}},\
  }\bibfield  {title} {\enquote {\bibinfo {title} {Preparation of regular
  arrays of antidots in {YBa$_2$Cu$_3$O$_7$} thin films and observation of
  vortex lattice matching effects},}\ }\href {\doibase 10.1063/1.119701}
  {\bibfield  {journal} {\bibinfo  {journal} {Appl. Phys. Lett.}\ }\textbf
  {\bibinfo {volume} {71}},\ \bibinfo {pages} {962} (\bibinfo {year}
  {1997})}\BibitemShut {NoStop}%
\bibitem [{\citenamefont {Lykov}(2011)}]{LYKO11}%
  \BibitemOpen
  \bibfield  {author} {\bibinfo {author} {\bibfnamefont {A.~N.}\ \bibnamefont
  {Lykov}},\ }\bibfield  {title} {\enquote {\bibinfo {title} {Unusual
  commensurability effect in superconducting {Sn} films with triangular lattice
  of microholes},}\ }\href {\doibase 10.1007/s10909-011-0363-z} {\bibfield
  {journal} {\bibinfo  {journal} {J. Low Temp. Phys.}\ }\textbf {\bibinfo
  {volume} {164}},\ \bibinfo {pages} {61} (\bibinfo {year}
  {2011})}\BibitemShut {NoStop}%
\bibitem [{\citenamefont {Crisan}\ \emph {et~al.}(2005)\citenamefont {Crisan},
  \citenamefont {Pross}, \citenamefont {Cole}, \citenamefont {Bending},
  \citenamefont {W\"ordenweber}, \citenamefont {Lahl},\ and\ \citenamefont
  {Brandt}}]{CRIS05}%
  \BibitemOpen
  \bibfield  {author} {\bibinfo {author} {\bibfnamefont {A.}~\bibnamefont
  {Crisan}}, \bibinfo {author} {\bibfnamefont {A.}~\bibnamefont {Pross}},
  \bibinfo {author} {\bibfnamefont {D.}~\bibnamefont {Cole}}, \bibinfo {author}
  {\bibfnamefont {S.~J.}\ \bibnamefont {Bending}}, \bibinfo {author}
  {\bibfnamefont {R.}~\bibnamefont {W\"ordenweber}}, \bibinfo {author}
  {\bibfnamefont {P.}~\bibnamefont {Lahl}}, \ and\ \bibinfo {author}
  {\bibfnamefont {E.~H.}\ \bibnamefont {Brandt}},\ }\bibfield  {title}
  {\enquote {\bibinfo {title} {Anisotropic vortex channeling in
  {YBa$_2$Cu$_3$O$_{7-\delta}$} thin films with ordered antidot arrays},}\
  }\href {\doibase 10.1103/physrevb.71.144504} {\bibfield  {journal} {\bibinfo
  {journal} {Phys. Rev. B}\ }\textbf {\bibinfo {volume} {71}},\ \bibinfo
  {pages} {144504} (\bibinfo {year} {2005})}\BibitemShut {NoStop}%
\bibitem [{\citenamefont {Ooi}\ \emph {et~al.}(2005)\citenamefont {Ooi},
  \citenamefont {Mochiku}, \citenamefont {Yu}, \citenamefont {Sadki},\ and\
  \citenamefont {Hirata}}]{OOI05}%
  \BibitemOpen
  \bibfield  {author} {\bibinfo {author} {\bibfnamefont {S.}~\bibnamefont
  {Ooi}}, \bibinfo {author} {\bibfnamefont {T.}~\bibnamefont {Mochiku}},
  \bibinfo {author} {\bibfnamefont {S.}~\bibnamefont {Yu}}, \bibinfo {author}
  {\bibfnamefont {E.~S.}\ \bibnamefont {Sadki}}, \ and\ \bibinfo {author}
  {\bibfnamefont {K.}~\bibnamefont {Hirata}},\ }\bibfield  {title} {\enquote
  {\bibinfo {title} {Matching effect of vortex lattice in
  {Bi$_2$Sr$_2$CaCu$_2$O$_{8+y}$} with artificial periodic defects},}\ }\href
  {\doibase 10.1016/j.physc.2005.02.139} {\bibfield  {journal} {\bibinfo
  {journal} {Physica C}\ }\textbf {\bibinfo {volume} {426}},\ \bibinfo {pages}
  {113} (\bibinfo {year} {2005})}\BibitemShut {NoStop}%
\bibitem [{\citenamefont {Avci}\ \emph {et~al.}(2010)\citenamefont {Avci},
  \citenamefont {Xiao}, \citenamefont {Hua}, \citenamefont {Imre},
  \citenamefont {Divan}, \citenamefont {Pearson}, \citenamefont {Welp},
  \citenamefont {Kwok},\ and\ \citenamefont {Crabtree}}]{AVCI10}%
  \BibitemOpen
  \bibfield  {author} {\bibinfo {author} {\bibfnamefont {S.}~\bibnamefont
  {Avci}}, \bibinfo {author} {\bibfnamefont {Z.~L.}\ \bibnamefont {Xiao}},
  \bibinfo {author} {\bibfnamefont {J.}~\bibnamefont {Hua}}, \bibinfo {author}
  {\bibfnamefont {A.}~\bibnamefont {Imre}}, \bibinfo {author} {\bibfnamefont
  {R.}~\bibnamefont {Divan}}, \bibinfo {author} {\bibfnamefont
  {J.}~\bibnamefont {Pearson}}, \bibinfo {author} {\bibfnamefont
  {U.}~\bibnamefont {Welp}}, \bibinfo {author} {\bibfnamefont {W.~K.}\
  \bibnamefont {Kwok}}, \ and\ \bibinfo {author} {\bibfnamefont {G.~W.}\
  \bibnamefont {Crabtree}},\ }\bibfield  {title} {\enquote {\bibinfo {title}
  {Matching effect and dynamic phases of vortex matter in
  {Bi$_2$Sr$_2$CaCu$_2$O$_8$} nanoribbon with a periodic array of holes},}\
  }\href {\doibase 10.1063/1.3473783} {\bibfield  {journal} {\bibinfo
  {journal} {Appl. Phys. Lett.}\ }\textbf {\bibinfo {volume} {97}},\ \bibinfo
  {pages} {042511} (\bibinfo {year} {2010})}\BibitemShut {NoStop}%
\bibitem [{\citenamefont {Silhanek}\ \emph {et~al.}(2011)\citenamefont
  {Silhanek}, \citenamefont {Gutierrez}, \citenamefont {Kramer}, \citenamefont
  {Ataklti}, \citenamefont {Van~de Vondel}, \citenamefont {Moshchalkov},\ and\
  \citenamefont {Sanchez}}]{SILH11}%
  \BibitemOpen
  \bibfield  {author} {\bibinfo {author} {\bibfnamefont {A.~V.}\ \bibnamefont
  {Silhanek}}, \bibinfo {author} {\bibfnamefont {J.}~\bibnamefont {Gutierrez}},
  \bibinfo {author} {\bibfnamefont {R.~B.~G.}\ \bibnamefont {Kramer}}, \bibinfo
  {author} {\bibfnamefont {G.~W.}\ \bibnamefont {Ataklti}}, \bibinfo {author}
  {\bibfnamefont {J.}~\bibnamefont {Van~de Vondel}}, \bibinfo {author}
  {\bibfnamefont {V.~V.}\ \bibnamefont {Moshchalkov}}, \ and\ \bibinfo {author}
  {\bibfnamefont {A.}~\bibnamefont {Sanchez}},\ }\bibfield  {title} {\enquote
  {\bibinfo {title} {Microscopic picture of the critical state in a
  superconductor with a periodic array of antidots},}\ }\href {\doibase
  10.1103/physrevb.83.024509} {\bibfield  {journal} {\bibinfo  {journal} {Phys.
  Rev. B}\ }\textbf {\bibinfo {volume} {83}},\ \bibinfo {pages} {024509}
  (\bibinfo {year} {2011})}\BibitemShut {NoStop}%
\bibitem [{\citenamefont {Sochnikov}\ \emph {et~al.}(2011)\citenamefont
  {Sochnikov}, \citenamefont {Bozovic}, \citenamefont {Shaulov},\ and\
  \citenamefont {Yeshurun}}]{SOCH11}%
  \BibitemOpen
  \bibfield  {author} {\bibinfo {author} {\bibfnamefont {I.}~\bibnamefont
  {Sochnikov}}, \bibinfo {author} {\bibfnamefont {I.}~\bibnamefont {Bozovic}},
  \bibinfo {author} {\bibfnamefont {A.}~\bibnamefont {Shaulov}}, \ and\
  \bibinfo {author} {\bibfnamefont {Y.}~\bibnamefont {Yeshurun}},\ }\bibfield
  {title} {\enquote {\bibinfo {title} {Uncorrelated behavior of fluxoids in
  superconducting double networks},}\ }\href {\doibase
  10.1103/physrevb.84.094530} {\bibfield  {journal} {\bibinfo  {journal} {Phys.
  Rev. B}\ }\textbf {\bibinfo {volume} {84}},\ \bibinfo {pages} {094530}
  (\bibinfo {year} {2011})}\BibitemShut {NoStop}%
\bibitem [{\citenamefont {Swiecicki}\ \emph {et~al.}(2012)\citenamefont
  {Swiecicki}, \citenamefont {Ulysse}, \citenamefont {Wolf}, \citenamefont
  {Bernard}, \citenamefont {Bergeal}, \citenamefont {Briatico}, \citenamefont
  {Faini}, \citenamefont {Lesueur},\ and\ \citenamefont {Villegas}}]{SWIE12}%
  \BibitemOpen
  \bibfield  {author} {\bibinfo {author} {\bibfnamefont {I.}~\bibnamefont
  {Swiecicki}}, \bibinfo {author} {\bibfnamefont {C.}~\bibnamefont {Ulysse}},
  \bibinfo {author} {\bibfnamefont {T.}~\bibnamefont {Wolf}}, \bibinfo {author}
  {\bibfnamefont {R.}~\bibnamefont {Bernard}}, \bibinfo {author} {\bibfnamefont
  {N.}~\bibnamefont {Bergeal}}, \bibinfo {author} {\bibfnamefont
  {J.}~\bibnamefont {Briatico}}, \bibinfo {author} {\bibfnamefont
  {G.}~\bibnamefont {Faini}}, \bibinfo {author} {\bibfnamefont
  {J.}~\bibnamefont {Lesueur}}, \ and\ \bibinfo {author} {\bibfnamefont
  {J.~E.}\ \bibnamefont {Villegas}},\ }\bibfield  {title} {\enquote {\bibinfo
  {title} {Strong field-matching effects in superconducting
  {YBa$_2$Cu$_3$O$_{7-\delta}$} films with vortex energy landscapes engineered
  via masked ion irradiation},}\ }\href {\doibase 10.1103/physrevb.85.224502}
  {\bibfield  {journal} {\bibinfo  {journal} {Phys. Rev. B}\ }\textbf {\bibinfo
  {volume} {85}},\ \bibinfo {pages} {224502} (\bibinfo {year}
  {2012})}\BibitemShut {NoStop}%
\bibitem [{\citenamefont {Shaw}\ \emph {et~al.}(2012)\citenamefont {Shaw},
  \citenamefont {Bag}, \citenamefont {Banerjee}, \citenamefont {Suderow},\ and\
  \citenamefont {Tamegai}}]{SHAW12}%
  \BibitemOpen
  \bibfield  {author} {\bibinfo {author} {\bibfnamefont {G.}~\bibnamefont
  {Shaw}}, \bibinfo {author} {\bibfnamefont {B.}~\bibnamefont {Bag}}, \bibinfo
  {author} {\bibfnamefont {S.~S.}\ \bibnamefont {Banerjee}}, \bibinfo {author}
  {\bibfnamefont {H.}~\bibnamefont {Suderow}}, \ and\ \bibinfo {author}
  {\bibfnamefont {T.}~\bibnamefont {Tamegai}},\ }\bibfield  {title} {\enquote
  {\bibinfo {title} {Generating strong magnetic flux shielding regions in a
  single crystal of {Bi$_2$Sr$_2$CaCu$_2$O$_8$} using a blind hole array},}\
  }\href {\doibase 10.1088/0953-2048/25/9/095016} {\bibfield  {journal}
  {\bibinfo  {journal} {Supercond. Sci. Technol.}\ }\textbf {\bibinfo {volume}
  {25}},\ \bibinfo {pages} {095016} (\bibinfo {year} {2012})}\BibitemShut
  {NoStop}%
\bibitem [{\citenamefont {Trastoy}\ \emph {et~al.}(2013)\citenamefont
  {Trastoy}, \citenamefont {Rouco}, \citenamefont {Ulysse}, \citenamefont
  {Bernard}, \citenamefont {Palau}, \citenamefont {Puig}, \citenamefont
  {Faini}, \citenamefont {Lesueur}, \citenamefont {Briatico},\ and\
  \citenamefont {Villegas}}]{TRAS13}%
  \BibitemOpen
  \bibfield  {author} {\bibinfo {author} {\bibfnamefont {J.}~\bibnamefont
  {Trastoy}}, \bibinfo {author} {\bibfnamefont {V.}~\bibnamefont {Rouco}},
  \bibinfo {author} {\bibfnamefont {C.}~\bibnamefont {Ulysse}}, \bibinfo
  {author} {\bibfnamefont {R.}~\bibnamefont {Bernard}}, \bibinfo {author}
  {\bibfnamefont {A.}~\bibnamefont {Palau}}, \bibinfo {author} {\bibfnamefont
  {T.}~\bibnamefont {Puig}}, \bibinfo {author} {\bibfnamefont {G.}~\bibnamefont
  {Faini}}, \bibinfo {author} {\bibfnamefont {J.}~\bibnamefont {Lesueur}},
  \bibinfo {author} {\bibfnamefont {J.}~\bibnamefont {Briatico}}, \ and\
  \bibinfo {author} {\bibfnamefont {J.~E.}\ \bibnamefont {Villegas}},\
  }\bibfield  {title} {\enquote {\bibinfo {title} {Unusual magneto-transport of
  {YBa$_2$Cu$_3$O$_{7-\delta}$} films due to the interplay of anisotropy,
  random disorder and nanoscale periodic pinning},}\ }\href {\doibase
  10.1088/1367-2630/15/10/103022} {\bibfield  {journal} {\bibinfo  {journal}
  {New J. Phys.}\ }\textbf {\bibinfo {volume} {15}},\ \bibinfo {pages} {103022}
  (\bibinfo {year} {2013})}\BibitemShut {NoStop}%
\bibitem [{\citenamefont {Haag}\ \emph {et~al.}(2014)\citenamefont {Haag},
  \citenamefont {Zechner}, \citenamefont {Lang}, \citenamefont {Dosmailov},
  \citenamefont {Bodea},\ and\ \citenamefont {Pedarnig}}]{HAAG14}%
  \BibitemOpen
  \bibfield  {author} {\bibinfo {author} {\bibfnamefont {L.~T.}\ \bibnamefont
  {Haag}}, \bibinfo {author} {\bibfnamefont {G.}~\bibnamefont {Zechner}},
  \bibinfo {author} {\bibfnamefont {W.}~\bibnamefont {Lang}}, \bibinfo {author}
  {\bibfnamefont {M.}~\bibnamefont {Dosmailov}}, \bibinfo {author}
  {\bibfnamefont {M.~A.}\ \bibnamefont {Bodea}}, \ and\ \bibinfo {author}
  {\bibfnamefont {J.~D.}\ \bibnamefont {Pedarnig}},\ }\bibfield  {title}
  {\enquote {\bibinfo {title} {Strong vortex matching effects in {YBCO} films
  with periodic modulations of the superconducting order parameter fabricated
  by masked ion irradiation},}\ }\href {\doibase 10.1016/j.physc.2014.03.032}
  {\bibfield  {journal} {\bibinfo  {journal} {Physica C}\ }\textbf {\bibinfo
  {volume} {503}},\ \bibinfo {pages} {75} (\bibinfo {year}
  {2014})}\BibitemShut {NoStop}%
\bibitem [{\citenamefont {Trastoy}\ \emph {et~al.}(2014)\citenamefont
  {Trastoy}, \citenamefont {Malnou}, \citenamefont {Ulysse}, \citenamefont
  {Bernard}, \citenamefont {Bergeal}, \citenamefont {Faini}, \citenamefont
  {Lesueur}, \citenamefont {Briatico},\ and\ \citenamefont
  {Villegas}}]{TRAS14}%
  \BibitemOpen
  \bibfield  {author} {\bibinfo {author} {\bibfnamefont {J.}~\bibnamefont
  {Trastoy}}, \bibinfo {author} {\bibfnamefont {M.}~\bibnamefont {Malnou}},
  \bibinfo {author} {\bibfnamefont {C.}~\bibnamefont {Ulysse}}, \bibinfo
  {author} {\bibfnamefont {R.}~\bibnamefont {Bernard}}, \bibinfo {author}
  {\bibfnamefont {N.}~\bibnamefont {Bergeal}}, \bibinfo {author} {\bibfnamefont
  {G.}~\bibnamefont {Faini}}, \bibinfo {author} {\bibfnamefont
  {J.}~\bibnamefont {Lesueur}}, \bibinfo {author} {\bibfnamefont
  {J.}~\bibnamefont {Briatico}}, \ and\ \bibinfo {author} {\bibfnamefont
  {J.~E.}\ \bibnamefont {Villegas}},\ }\bibfield  {title} {\enquote {\bibinfo
  {title} {Freezing and thawing of artificial ice by thermal switching of
  geometric frustration in magnetic flux lattices},}\ }\href {\doibase
  10.1038/nnano.2014.158} {\bibfield  {journal} {\bibinfo  {journal} {Nat.
  Nanotechnol.}\ }\textbf {\bibinfo {volume} {9}},\ \bibinfo {pages} {710}
  (\bibinfo {year} {2014})}\BibitemShut {NoStop}%
\bibitem [{\citenamefont {Trastoy}\ \emph {et~al.}(2015)\citenamefont
  {Trastoy}, \citenamefont {Ulysse}, \citenamefont {Bernard}, \citenamefont
  {Malnou}, \citenamefont {Bergeal}, \citenamefont {Lesueur}, \citenamefont
  {Briatico},\ and\ \citenamefont {Villegas}}]{TRAS15}%
  \BibitemOpen
  \bibfield  {author} {\bibinfo {author} {\bibfnamefont {J.}~\bibnamefont
  {Trastoy}}, \bibinfo {author} {\bibfnamefont {C.}~\bibnamefont {Ulysse}},
  \bibinfo {author} {\bibfnamefont {R.}~\bibnamefont {Bernard}}, \bibinfo
  {author} {\bibfnamefont {M.}~\bibnamefont {Malnou}}, \bibinfo {author}
  {\bibfnamefont {N.}~\bibnamefont {Bergeal}}, \bibinfo {author} {\bibfnamefont
  {J.}~\bibnamefont {Lesueur}}, \bibinfo {author} {\bibfnamefont
  {J.}~\bibnamefont {Briatico}}, \ and\ \bibinfo {author} {\bibfnamefont
  {J.~E.}\ \bibnamefont {Villegas}},\ }\bibfield  {title} {\enquote {\bibinfo
  {title} {Tunable flux-matching effects in high-{$T_c$} superconductors with
  nonuniform pinning arrays},}\ }\href {\doibase
  10.1103/physrevapplied.4.054003} {\bibfield  {journal} {\bibinfo  {journal}
  {Phys. Rev. Appl.}\ }\textbf {\bibinfo {volume} {4}},\ \bibinfo {pages}
  {054003} (\bibinfo {year} {2015})}\BibitemShut {NoStop}%
\bibitem [{\citenamefont {Poccia}\ \emph {et~al.}(2015)\citenamefont {Poccia},
  \citenamefont {Baturina}, \citenamefont {Coneri}, \citenamefont {Molenaar},
  \citenamefont {Wang}, \citenamefont {Bianconi}, \citenamefont {Brinkman},
  \citenamefont {Hilgenkamp}, \citenamefont {Golubov},\ and\ \citenamefont
  {Vinokur}}]{POCC15}%
  \BibitemOpen
  \bibfield  {author} {\bibinfo {author} {\bibfnamefont {N.}~\bibnamefont
  {Poccia}}, \bibinfo {author} {\bibfnamefont {T.~I.}\ \bibnamefont
  {Baturina}}, \bibinfo {author} {\bibfnamefont {F.}~\bibnamefont {Coneri}},
  \bibinfo {author} {\bibfnamefont {C.~G.}\ \bibnamefont {Molenaar}}, \bibinfo
  {author} {\bibfnamefont {X.~R.}\ \bibnamefont {Wang}}, \bibinfo {author}
  {\bibfnamefont {G.}~\bibnamefont {Bianconi}}, \bibinfo {author}
  {\bibfnamefont {A.}~\bibnamefont {Brinkman}}, \bibinfo {author}
  {\bibfnamefont {H.}~\bibnamefont {Hilgenkamp}}, \bibinfo {author}
  {\bibfnamefont {A.~A.}\ \bibnamefont {Golubov}}, \ and\ \bibinfo {author}
  {\bibfnamefont {V.~M.}\ \bibnamefont {Vinokur}},\ }\bibfield  {title}
  {\enquote {\bibinfo {title} {Critical behavior at a dynamic vortex
  insulator-to-metal transition},}\ }\href {\doibase 10.1126/science.1260507}
  {\bibfield  {journal} {\bibinfo  {journal} {Science}\ }\textbf {\bibinfo
  {volume} {349}},\ \bibinfo {pages} {1202} (\bibinfo {year}
  {2015})}\BibitemShut {NoStop}%
\bibitem [{\citenamefont {W\"ordenweber}\ \emph {et~al.}(2004)\citenamefont
  {W\"ordenweber}, \citenamefont {Dymashevski},\ and\ \citenamefont
  {Misko}}]{WORD04}%
  \BibitemOpen
  \bibfield  {author} {\bibinfo {author} {\bibfnamefont {R.}~\bibnamefont
  {W\"ordenweber}}, \bibinfo {author} {\bibfnamefont {P.}~\bibnamefont
  {Dymashevski}}, \ and\ \bibinfo {author} {\bibfnamefont {V.~R.}\ \bibnamefont
  {Misko}},\ }\bibfield  {title} {\enquote {\bibinfo {title} {Guidance of
  vortices and the vortex ratchet effect in high-{$T_c$} superconducting thin
  films obtained by arrangement of antidots},}\ }\href {\doibase
  10.1103/physrevb.69.184504} {\bibfield  {journal} {\bibinfo  {journal} {Phys.
  Rev. B}\ }\textbf {\bibinfo {volume} {69}},\ \bibinfo {pages} {184504}
  (\bibinfo {year} {2004})}\BibitemShut {NoStop}%
\bibitem [{\citenamefont {Ooi}\ \emph {et~al.}(2007)\citenamefont {Ooi},
  \citenamefont {Savel'ev}, \citenamefont {Gaifullin}, \citenamefont {Mochiku},
  \citenamefont {Hirata},\ and\ \citenamefont {Nori}}]{OOI07}%
  \BibitemOpen
  \bibfield  {author} {\bibinfo {author} {\bibfnamefont {S.}~\bibnamefont
  {Ooi}}, \bibinfo {author} {\bibfnamefont {S.}~\bibnamefont {Savel'ev}},
  \bibinfo {author} {\bibfnamefont {M.~B.}\ \bibnamefont {Gaifullin}}, \bibinfo
  {author} {\bibfnamefont {T.}~\bibnamefont {Mochiku}}, \bibinfo {author}
  {\bibfnamefont {K.}~\bibnamefont {Hirata}}, \ and\ \bibinfo {author}
  {\bibfnamefont {F.}~\bibnamefont {Nori}},\ }\bibfield  {title} {\enquote
  {\bibinfo {title} {Nonlinear nanodevices using magnetic flux quanta},}\
  }\href {\doibase 10.1103/physrevlett.99.207003} {\bibfield  {journal}
  {\bibinfo  {journal} {Phys. Rev. Lett.}\ }\textbf {\bibinfo {volume} {99}},\
  \bibinfo {pages} {207003} (\bibinfo {year} {2007})}\BibitemShut {NoStop}%
\bibitem [{\citenamefont {Palau}\ \emph {et~al.}(2012)\citenamefont {Palau},
  \citenamefont {Monton}, \citenamefont {Rouco}, \citenamefont {Obradors},\
  and\ \citenamefont {Puig}}]{PALA12}%
  \BibitemOpen
  \bibfield  {author} {\bibinfo {author} {\bibfnamefont {A.}~\bibnamefont
  {Palau}}, \bibinfo {author} {\bibfnamefont {C.}~\bibnamefont {Monton}},
  \bibinfo {author} {\bibfnamefont {V.}~\bibnamefont {Rouco}}, \bibinfo
  {author} {\bibfnamefont {X.}~\bibnamefont {Obradors}}, \ and\ \bibinfo
  {author} {\bibfnamefont {T.}~\bibnamefont {Puig}},\ }\bibfield  {title}
  {\enquote {\bibinfo {title} {Guided vortex motion in {YBa$_2$Cu$_3$O$_7$}
  thin films with collective ratchet pinning potentials},}\ }\href {\doibase
  10.1103/physrevb.85.012502} {\bibfield  {journal} {\bibinfo  {journal} {Phys.
  Rev. B}\ }\textbf {\bibinfo {volume} {85}},\ \bibinfo {pages} {012502}
  (\bibinfo {year} {2012})}\BibitemShut {NoStop}%
\bibitem [{\citenamefont {Hastings}\ \emph {et~al.}(2003)\citenamefont
  {Hastings}, \citenamefont {Olson-Reichhardt},\ and\ \citenamefont
  {Reichhardt}}]{HAST03}%
  \BibitemOpen
  \bibfield  {author} {\bibinfo {author} {\bibfnamefont {M.~B.}\ \bibnamefont
  {Hastings}}, \bibinfo {author} {\bibfnamefont {C.~J.}\ \bibnamefont
  {Olson-Reichhardt}}, \ and\ \bibinfo {author} {\bibfnamefont
  {C.}~\bibnamefont {Reichhardt}},\ }\bibfield  {title} {\enquote {\bibinfo
  {title} {Ratchet cellular automata},}\ }\href {\doibase
  10.1103/physrevlett.90.247004} {\bibfield  {journal} {\bibinfo  {journal}
  {Phys. Rev. Lett.}\ }\textbf {\bibinfo {volume} {90}},\ \bibinfo {pages}
  {247004} (\bibinfo {year} {2003})}\BibitemShut {NoStop}%
\bibitem [{\citenamefont {Milo{\v{s}}evi{\'c}}\ \emph
  {et~al.}(2007)\citenamefont {Milo{\v{s}}evi{\'c}}, \citenamefont
  {Berdiyorov},\ and\ \citenamefont {Peeters}}]{MILO07}%
  \BibitemOpen
  \bibfield  {author} {\bibinfo {author} {\bibfnamefont {M.~V.}\ \bibnamefont
  {Milo{\v{s}}evi{\'c}}}, \bibinfo {author} {\bibfnamefont {G.~R.}\
  \bibnamefont {Berdiyorov}}, \ and\ \bibinfo {author} {\bibfnamefont {F.~M.}\
  \bibnamefont {Peeters}},\ }\bibfield  {title} {\enquote {\bibinfo {title}
  {Fluxonic cellular automata},}\ }\href {\doibase 10.1063/1.2813047}
  {\bibfield  {journal} {\bibinfo  {journal} {Appl. Phys. Lett.}\ }\textbf
  {\bibinfo {volume} {91}},\ \bibinfo {pages} {212501} (\bibinfo {year}
  {2007})}\BibitemShut {NoStop}%
\bibitem [{\citenamefont {Dam}\ \emph {et~al.}(1999)\citenamefont {Dam},
  \citenamefont {Huijbregtse}, \citenamefont {Klaassen}, \citenamefont {van-der
  Geest}, \citenamefont {Doornbos}, \citenamefont {Rector}, \citenamefont
  {Testa}, \citenamefont {Freisem}, \citenamefont {Martinez}, \citenamefont
  {Stauble-Pumpin},\ and\ \citenamefont {Griessen}}]{DAM99}%
  \BibitemOpen
  \bibfield  {author} {\bibinfo {author} {\bibfnamefont {B.}~\bibnamefont
  {Dam}}, \bibinfo {author} {\bibfnamefont {J.~M.}\ \bibnamefont
  {Huijbregtse}}, \bibinfo {author} {\bibfnamefont {F.~C.}\ \bibnamefont
  {Klaassen}}, \bibinfo {author} {\bibfnamefont {R.~C.~F.}\ \bibnamefont
  {van-der Geest}}, \bibinfo {author} {\bibfnamefont {G.}~\bibnamefont
  {Doornbos}}, \bibinfo {author} {\bibfnamefont {J.~H.}\ \bibnamefont
  {Rector}}, \bibinfo {author} {\bibfnamefont {A.~M.}\ \bibnamefont {Testa}},
  \bibinfo {author} {\bibfnamefont {S.}~\bibnamefont {Freisem}}, \bibinfo
  {author} {\bibfnamefont {J.~C.}\ \bibnamefont {Martinez}}, \bibinfo {author}
  {\bibfnamefont {B.}~\bibnamefont {Stauble-Pumpin}}, \ and\ \bibinfo {author}
  {\bibfnamefont {R.}~\bibnamefont {Griessen}},\ }\bibfield  {title} {\enquote
  {\bibinfo {title} {Origin of high critical currents in {YBa$_2$Cu$_3$O$_{7-
  \delta}$} superconducting thin films},}\ }\href {\doibase 10.1038/20880}
  {\bibfield  {journal} {\bibinfo  {journal} {Nature}\ }\textbf {\bibinfo
  {volume} {399}},\ \bibinfo {pages} {439} (\bibinfo {year}
  {1999})}\BibitemShut {NoStop}%
\bibitem [{\citenamefont {Lang}\ \emph {et~al.}(2006)\citenamefont {Lang},
  \citenamefont {Dineva}, \citenamefont {Marksteiner}, \citenamefont
  {Enzenhofer}, \citenamefont {Siraj}, \citenamefont {Peruzzi}, \citenamefont
  {Pedarnig}, \citenamefont {B\"auerle}, \citenamefont {Korntner},
  \citenamefont {Cekan}, \citenamefont {Platzgummer},\ and\ \citenamefont
  {Loeschner}}]{LANG06a}%
  \BibitemOpen
  \bibfield  {author} {\bibinfo {author} {\bibfnamefont {W.}~\bibnamefont
  {Lang}}, \bibinfo {author} {\bibfnamefont {M.}~\bibnamefont {Dineva}},
  \bibinfo {author} {\bibfnamefont {M.}~\bibnamefont {Marksteiner}}, \bibinfo
  {author} {\bibfnamefont {T.}~\bibnamefont {Enzenhofer}}, \bibinfo {author}
  {\bibfnamefont {K.}~\bibnamefont {Siraj}}, \bibinfo {author} {\bibfnamefont
  {M.}~\bibnamefont {Peruzzi}}, \bibinfo {author} {\bibfnamefont {J.~D.}\
  \bibnamefont {Pedarnig}}, \bibinfo {author} {\bibfnamefont {D.}~\bibnamefont
  {B\"auerle}}, \bibinfo {author} {\bibfnamefont {R.}~\bibnamefont {Korntner}},
  \bibinfo {author} {\bibfnamefont {E.}~\bibnamefont {Cekan}}, \bibinfo
  {author} {\bibfnamefont {E.}~\bibnamefont {Platzgummer}}, \ and\ \bibinfo
  {author} {\bibfnamefont {H.}~\bibnamefont {Loeschner}},\ }\bibfield  {title}
  {\enquote {\bibinfo {title} {Ion-beam direct-structuring of high-temperature
  superconductors},}\ }\href {\doibase 10.1016/j.mee.2006.01.091} {\bibfield
  {journal} {\bibinfo  {journal} {Microelectron. Eng.}\ }\textbf {\bibinfo
  {volume} {83}},\ \bibinfo {pages} {1495} (\bibinfo {year}
  {2006})}\BibitemShut {NoStop}%
\bibitem [{\citenamefont {Pedarnig}\ \emph {et~al.}(2010)\citenamefont
  {Pedarnig}, \citenamefont {Siraj}, \citenamefont {Bodea}, \citenamefont
  {Puica}, \citenamefont {Lang}, \citenamefont {Kolarova}, \citenamefont
  {Bauer}, \citenamefont {Haselgr\"ubler}, \citenamefont {Hasenfuss},
  \citenamefont {Beinik},\ and\ \citenamefont {Teichert}}]{PEDA10}%
  \BibitemOpen
  \bibfield  {author} {\bibinfo {author} {\bibfnamefont {J.~D.}\ \bibnamefont
  {Pedarnig}}, \bibinfo {author} {\bibfnamefont {K.}~\bibnamefont {Siraj}},
  \bibinfo {author} {\bibfnamefont {M.~A.}\ \bibnamefont {Bodea}}, \bibinfo
  {author} {\bibfnamefont {I.}~\bibnamefont {Puica}}, \bibinfo {author}
  {\bibfnamefont {W.}~\bibnamefont {Lang}}, \bibinfo {author} {\bibfnamefont
  {R.}~\bibnamefont {Kolarova}}, \bibinfo {author} {\bibfnamefont
  {P.}~\bibnamefont {Bauer}}, \bibinfo {author} {\bibfnamefont
  {K.}~\bibnamefont {Haselgr\"ubler}}, \bibinfo {author} {\bibfnamefont
  {C.}~\bibnamefont {Hasenfuss}}, \bibinfo {author} {\bibfnamefont
  {I.}~\bibnamefont {Beinik}}, \ and\ \bibinfo {author} {\bibfnamefont
  {C.}~\bibnamefont {Teichert}},\ }\bibfield  {title} {\enquote {\bibinfo
  {title} {Surface planarization and masked ion-beam structuring of
  {YBa$_2$Cu$_3$O$_7$} thin films},}\ }\href {\doibase
  10.1016/j.tsf.2010.07.021} {\bibfield  {journal} {\bibinfo  {journal} {Thin
  Solid Films}\ }\textbf {\bibinfo {volume} {518}},\ \bibinfo {pages}
  {7075} (\bibinfo {year} {2010})}\BibitemShut {NoStop}%
\bibitem [{\citenamefont {Kahlmann}\ \emph {et~al.}(1998)\citenamefont
  {Kahlmann}, \citenamefont {Engelhardt}, \citenamefont {Schubert},
  \citenamefont {Zander}, \citenamefont {Buchal},\ and\ \citenamefont
  {Hollkott}}]{KAHL98}%
  \BibitemOpen
  \bibfield  {author} {\bibinfo {author} {\bibfnamefont {F.}~\bibnamefont
  {Kahlmann}}, \bibinfo {author} {\bibfnamefont {A.}~\bibnamefont
  {Engelhardt}}, \bibinfo {author} {\bibfnamefont {J.}~\bibnamefont
  {Schubert}}, \bibinfo {author} {\bibfnamefont {W.}~\bibnamefont {Zander}},
  \bibinfo {author} {\bibfnamefont {Ch}~\bibnamefont {Buchal}}, \ and\ \bibinfo
  {author} {\bibfnamefont {J.}~\bibnamefont {Hollkott}},\ }\bibfield  {title}
  {\enquote {\bibinfo {title} {Superconductor-normal-superconductor {J}osephson
  junctions fabricated by oxygen implantation into
  {YBa$_2$Cu$_3$O$_{7-\delta}$}},}\ }\href {\doibase 10.1063/1.122459}
  {\bibfield  {journal} {\bibinfo  {journal} {Appl. Phys. Lett.}\ }\textbf
  {\bibinfo {volume} {73}},\ \bibinfo {pages} {2354} (\bibinfo {year}
  {1998})}\BibitemShut {NoStop}%
\bibitem [{\citenamefont {Katz}\ \emph {et~al.}(2000)\citenamefont {Katz},
  \citenamefont {Woods},\ and\ \citenamefont {Dynes}}]{KATZ00}%
  \BibitemOpen
  \bibfield  {author} {\bibinfo {author} {\bibfnamefont {A.~S.}\ \bibnamefont
  {Katz}}, \bibinfo {author} {\bibfnamefont {S.~I.}\ \bibnamefont {Woods}}, \
  and\ \bibinfo {author} {\bibfnamefont {R.~C.}\ \bibnamefont {Dynes}},\
  }\bibfield  {title} {\enquote {\bibinfo {title} {Transport properties of
  high-{$T_c$} planar {J}osephson junctions fabricated by nanolithography and
  ion implantation},}\ }\href {\doibase 10.1063/1.372286} {\bibfield  {journal}
  {\bibinfo  {journal} {J. Appl. Phys.}\ }\textbf {\bibinfo {volume} {87}},\
  \bibinfo {pages} {2978} (\bibinfo {year} {2000})}\BibitemShut {NoStop}%
\bibitem [{\citenamefont {Kang}\ \emph {et~al.}(2002)\citenamefont {Kang},
  \citenamefont {Burnell}, \citenamefont {Lloyd}, \citenamefont {Speaks},
  \citenamefont {Peng}, \citenamefont {Jeynes}, \citenamefont {Webb},
  \citenamefont {Yun}, \citenamefont {Moon}, \citenamefont {Oh}, \citenamefont
  {Tarte}, \citenamefont {Moore},\ and\ \citenamefont {Blamire}}]{KANG02a}%
  \BibitemOpen
  \bibfield  {author} {\bibinfo {author} {\bibfnamefont {D.~J.}\ \bibnamefont
  {Kang}}, \bibinfo {author} {\bibfnamefont {G.}~\bibnamefont {Burnell}},
  \bibinfo {author} {\bibfnamefont {S.~J.}\ \bibnamefont {Lloyd}}, \bibinfo
  {author} {\bibfnamefont {R.~S.}\ \bibnamefont {Speaks}}, \bibinfo {author}
  {\bibfnamefont {N.~H.}\ \bibnamefont {Peng}}, \bibinfo {author}
  {\bibfnamefont {C.}~\bibnamefont {Jeynes}}, \bibinfo {author} {\bibfnamefont
  {R.}~\bibnamefont {Webb}}, \bibinfo {author} {\bibfnamefont {J.~H.}\
  \bibnamefont {Yun}}, \bibinfo {author} {\bibfnamefont {S.~H.}\ \bibnamefont
  {Moon}}, \bibinfo {author} {\bibfnamefont {B.}~\bibnamefont {Oh}}, \bibinfo
  {author} {\bibfnamefont {E.~J.}\ \bibnamefont {Tarte}}, \bibinfo {author}
  {\bibfnamefont {D.~F.}\ \bibnamefont {Moore}}, \ and\ \bibinfo {author}
  {\bibfnamefont {M.~G.}\ \bibnamefont {Blamire}},\ }\bibfield  {title}
  {\enquote {\bibinfo {title} {Realization and properties of
  {YBa$_2$Cu$_3$O$_{7-\delta}$} {J}osephson junctions by metal masked ion
  damage technique},}\ }\href {\doibase 10.1063/1.1446998} {\bibfield
  {journal} {\bibinfo  {journal} {Appl. Phys. Lett.}\ }\textbf {\bibinfo
  {volume} {80}},\ \bibinfo {pages} {814} (\bibinfo {year}
  {2002})}\BibitemShut {NoStop}%
\bibitem [{\citenamefont {Blamire}\ \emph {et~al.}(2003)\citenamefont
  {Blamire}, \citenamefont {Kang}, \citenamefont {Burnell}, \citenamefont
  {Peng}, \citenamefont {Webb}, \citenamefont {Jeynes}, \citenamefont {Yun},
  \citenamefont {Moon},\ and\ \citenamefont {Oh}}]{BLAM03}%
  \BibitemOpen
  \bibfield  {author} {\bibinfo {author} {\bibfnamefont {M.~G.}\ \bibnamefont
  {Blamire}}, \bibinfo {author} {\bibfnamefont {D.~J.}\ \bibnamefont {Kang}},
  \bibinfo {author} {\bibfnamefont {G.}~\bibnamefont {Burnell}}, \bibinfo
  {author} {\bibfnamefont {N.~H.}\ \bibnamefont {Peng}}, \bibinfo {author}
  {\bibfnamefont {R.}~\bibnamefont {Webb}}, \bibinfo {author} {\bibfnamefont
  {C.}~\bibnamefont {Jeynes}}, \bibinfo {author} {\bibfnamefont {J.~H.}\
  \bibnamefont {Yun}}, \bibinfo {author} {\bibfnamefont {S.~H.}\ \bibnamefont
  {Moon}}, \ and\ \bibinfo {author} {\bibfnamefont {B.}~\bibnamefont {Oh}},\
  }\bibfield  {title} {\enquote {\bibinfo {title} {Masked ion damage and
  implantation for device fabrication},}\ }\href {\doibase
  10.1016/s0042-207x(02)00303-2} {\bibfield  {journal} {\bibinfo  {journal}
  {Vacuum}\ }\textbf {\bibinfo {volume} {69}},\ \bibinfo {pages} {11}
  (\bibinfo {year} {2003})}\BibitemShut {NoStop}%
\bibitem [{\citenamefont {Bergeal}\ \emph {et~al.}(2005)\citenamefont
  {Bergeal}, \citenamefont {Grison}, \citenamefont {Lesueur}, \citenamefont
  {Faini}, \citenamefont {Aprili},\ and\ \citenamefont {Contour}}]{BERG05}%
  \BibitemOpen
  \bibfield  {author} {\bibinfo {author} {\bibfnamefont {N.}~\bibnamefont
  {Bergeal}}, \bibinfo {author} {\bibfnamefont {X.}~\bibnamefont {Grison}},
  \bibinfo {author} {\bibfnamefont {J.}~\bibnamefont {Lesueur}}, \bibinfo
  {author} {\bibfnamefont {G.}~\bibnamefont {Faini}}, \bibinfo {author}
  {\bibfnamefont {M.}~\bibnamefont {Aprili}}, \ and\ \bibinfo {author}
  {\bibfnamefont {J.~P.}\ \bibnamefont {Contour}},\ }\bibfield  {title}
  {\enquote {\bibinfo {title} {High-quality planar high-$t_c$ {J}osephson
  junctions},}\ }\href {\doibase 10.1063/1.2037206} {\bibfield  {journal}
  {\bibinfo  {journal} {Appl. Phys. Lett.}\ }\textbf {\bibinfo {volume} {87}},\
  \bibinfo {pages} {102502} (\bibinfo {year} {2005})}\BibitemShut {NoStop}%
\bibitem [{\citenamefont {Heine}\ and\ \citenamefont {Lang}(1998)}]{HEIN98}%
  \BibitemOpen
  \bibfield  {author} {\bibinfo {author} {\bibfnamefont {G.}~\bibnamefont
  {Heine}}\ and\ \bibinfo {author} {\bibfnamefont {W.}~\bibnamefont {Lang}},\
  }\bibfield  {title} {\enquote {\bibinfo {title} {Magnetoresistance of the new
  ceramic {`Cernox'} thermometer from {4.2~K} to {300~K} in magnetic fields up
  to {13~T}},}\ }\href {\doibase 10.1016/s0011-2275(97)00130-6} {\bibfield
  {journal} {\bibinfo  {journal} {Cryogenics}\ }\textbf {\bibinfo {volume}
  {38}},\ \bibinfo {pages} {377} (\bibinfo {year} {1998})}\BibitemShut
  {NoStop}%
\bibitem [{\citenamefont {Metlushko}\ \emph {et~al.}(1994)\citenamefont
  {Metlushko}, \citenamefont {Baert}, \citenamefont {Jonckheere}, \citenamefont
  {Moshchalkov},\ and\ \citenamefont {Bruynseraede}}]{METL94}%
  \BibitemOpen
  \bibfield  {author} {\bibinfo {author} {\bibfnamefont {V.~V.}\ \bibnamefont
  {Metlushko}}, \bibinfo {author} {\bibfnamefont {M.}~\bibnamefont {Baert}},
  \bibinfo {author} {\bibfnamefont {R.}~\bibnamefont {Jonckheere}}, \bibinfo
  {author} {\bibfnamefont {V.~V.}\ \bibnamefont {Moshchalkov}}, \ and\ \bibinfo
  {author} {\bibfnamefont {Y.}~\bibnamefont {Bruynseraede}},\ }\bibfield
  {title} {\enquote {\bibinfo {title} {Matching effects in {Pb/Ge} multilayers
  with the lattice of submicron holes},}\ }\href {\doibase
  10.1016/0038-1098(94)90628-9} {\bibfield  {journal} {\bibinfo  {journal}
  {Solid State Commun.}\ }\textbf {\bibinfo {volume} {91}},\ \bibinfo {pages}
  {331} (\bibinfo {year} {1994})}\BibitemShut {NoStop}%
\bibitem [{\citenamefont {Berdiyorov}\ \emph {et~al.}(2006)\citenamefont
  {Berdiyorov}, \citenamefont {Milo{\v{s}}evi{\'c}},\ and\ \citenamefont
  {Peeters}}]{BERD06}%
  \BibitemOpen
  \bibfield  {author} {\bibinfo {author} {\bibfnamefont {G.~R.}\ \bibnamefont
  {Berdiyorov}}, \bibinfo {author} {\bibfnamefont {M.~V.}\ \bibnamefont
  {Milo{\v{s}}evi{\'c}}}, \ and\ \bibinfo {author} {\bibfnamefont {F.~M.}\
  \bibnamefont {Peeters}},\ }\bibfield  {title} {\enquote {\bibinfo {title}
  {Vortex configurations and critical parameters in superconducting thin films
  containing antidot arrays: {N}onlinear {G}inzburg-{L}andau theory},}\ }\href
  {\doibase 10.1103/PhysRevB.74.174512} {\bibfield  {journal} {\bibinfo
  {journal} {Phys. Rev. B}\ }\textbf {\bibinfo {volume} {74}},\ \bibinfo
  {pages} {174512} (\bibinfo {year} {2006})}\BibitemShut {NoStop}%
\bibitem [{\citenamefont {Lee}\ and\ \citenamefont {Lemberger}(1993)}]{LEE93}%
  \BibitemOpen
  \bibfield  {author} {\bibinfo {author} {\bibfnamefont {J.-Y.}\ \bibnamefont
  {Lee}}\ and\ \bibinfo {author} {\bibfnamefont {T.~R.}\ \bibnamefont
  {Lemberger}},\ }\bibfield  {title} {\enquote {\bibinfo {title} {Penetration
  depth $\lambda$(t) of {YBa$_2$Cu$_3$O$_{7-\delta}$} films determined from the
  kinetic inductance},}\ }\href {\doibase 10.1063/1.109383} {\bibfield
  {journal} {\bibinfo  {journal} {Appl. Phys. Lett.}\ }\textbf {\bibinfo
  {volume} {62}},\ \bibinfo {pages} {2419} (\bibinfo {year}
  {1993})}\BibitemShut {NoStop}%
\bibitem [{\citenamefont {Brandt}\ and\ \citenamefont
  {Indenbom}(1993)}]{BRAN93c}%
  \BibitemOpen
  \bibfield  {author} {\bibinfo {author} {\bibfnamefont {E.~H.}\ \bibnamefont
  {Brandt}}\ and\ \bibinfo {author} {\bibfnamefont {M.}~\bibnamefont
  {Indenbom}},\ }\bibfield  {title} {\enquote {\bibinfo {title}
  {Type-{II}-superconductor strip with current in a perpendicular magnetic
  field},}\ }\href {\doibase 10.1103/PhysRevB.48.12893} {\bibfield  {journal}
  {\bibinfo  {journal} {Phys. Rev. B}\ }\textbf {\bibinfo {volume} {48}},\
  \bibinfo {pages} {12893} (\bibinfo {year} {1993})}\BibitemShut
  {NoStop}%
\bibitem [{\citenamefont {Lang}(1995)}]{LANG95e}%
  \BibitemOpen
  \bibfield  {author} {\bibinfo {author} {\bibfnamefont {W.}~\bibnamefont
  {Lang}},\ }\bibfield  {title} {\enquote {\bibinfo {title} {Study of
  superconducting fluctuations in the high-temperature superconductor
  {YBa$_2$Cu$_3$O$_7$} by magnetotransport measurements},}\ }\href {\doibase
  10.1016/0379-6779(94)02949-y} {\bibfield  {journal} {\bibinfo  {journal}
  {Synthetic Met.}\ }\textbf {\bibinfo {volume} {71}},\ \bibinfo {pages}
  {1555} (\bibinfo {year} {1995})}\BibitemShut {NoStop}%
\bibitem [{\citenamefont {Buzdin}(1993)}]{BUZD93}%
  \BibitemOpen
  \bibfield  {author} {\bibinfo {author} {\bibfnamefont {A.~I.}\ \bibnamefont
  {Buzdin}},\ }\bibfield  {title} {\enquote {\bibinfo {title} {Multiple-quanta
  vortices at columnar defects},}\ }\href {\doibase 10.1103/PhysRevB.47.11416}
  {\bibfield  {journal} {\bibinfo  {journal} {Phys. Rev. B}\ }\textbf {\bibinfo
  {volume} {47}},\ \bibinfo {pages} {11416} (\bibinfo {year}
  {1993})}\BibitemShut {NoStop}%
\bibitem [{\citenamefont {Evetts}\ and\ \citenamefont
  {Glowacki}(1988)}]{EVET88}%
  \BibitemOpen
  \bibfield  {author} {\bibinfo {author} {\bibfnamefont {J.~E.}\ \bibnamefont
  {Evetts}}\ and\ \bibinfo {author} {\bibfnamefont {B.~A.}\ \bibnamefont
  {Glowacki}},\ }\bibfield  {title} {\enquote {\bibinfo {title} {Relation of
  critical current irreversibility to trapped flux and microstructure in
  polycrystalline {YBa$_2$Cu$_3$O$_7$}},}\ }\href {\doibase
  10.1016/0011-2275(88)90147-6} {\bibfield  {journal} {\bibinfo  {journal}
  {Cryogenics}\ }\textbf {\bibinfo {volume} {28}},\ \bibinfo {pages} {641}
  (\bibinfo {year} {1988})}\BibitemShut {NoStop}%
\bibitem [{\citenamefont {Ji}\ \emph {et~al.}(1993)\citenamefont {Ji},
  \citenamefont {Rzchowski}, \citenamefont {Anand},\ and\ \citenamefont
  {Tinkham}}]{JI93}%
  \BibitemOpen
  \bibfield  {author} {\bibinfo {author} {\bibfnamefont {L.}~\bibnamefont
  {Ji}}, \bibinfo {author} {\bibfnamefont {M.~S.}\ \bibnamefont {Rzchowski}},
  \bibinfo {author} {\bibfnamefont {N.}~\bibnamefont {Anand}}, \ and\ \bibinfo
  {author} {\bibfnamefont {M.}~\bibnamefont {Tinkham}},\ }\bibfield  {title}
  {\enquote {\bibinfo {title} {Magnetic-field-dependent surface resistance and
  two-level critical-state model for granular superconductors},}\ }\href
  {\doibase 10.1103/PhysRevB.47.470} {\bibfield  {journal} {\bibinfo  {journal}
  {Phys. Rev. B}\ }\textbf {\bibinfo {volume} {47}},\ \bibinfo {pages}
  {470} (\bibinfo {year} {1993})}\BibitemShut {NoStop}%
\bibitem [{\citenamefont {Massarotti}\ \emph {et~al.}(2016)\citenamefont
  {Massarotti}, \citenamefont {Jouault}, \citenamefont {Rouco}, \citenamefont
  {Charpentier}, \citenamefont {Bauch}, \citenamefont {Michon}, \citenamefont
  {De~Candia}, \citenamefont {Lucignano}, \citenamefont {Lombardi},
  \citenamefont {Tafuri},\ and\ \citenamefont {Tagliacozzo}}]{MASS16}%
  \BibitemOpen
  \bibfield  {author} {\bibinfo {author} {\bibfnamefont {D.}~\bibnamefont
  {Massarotti}}, \bibinfo {author} {\bibfnamefont {B.}~\bibnamefont {Jouault}},
  \bibinfo {author} {\bibfnamefont {V.}~\bibnamefont {Rouco}}, \bibinfo
  {author} {\bibfnamefont {S.}~\bibnamefont {Charpentier}}, \bibinfo {author}
  {\bibfnamefont {T.}~\bibnamefont {Bauch}}, \bibinfo {author} {\bibfnamefont
  {A.}~\bibnamefont {Michon}}, \bibinfo {author} {\bibfnamefont
  {A.}~\bibnamefont {De~Candia}}, \bibinfo {author} {\bibfnamefont
  {P.}~\bibnamefont {Lucignano}}, \bibinfo {author} {\bibfnamefont
  {F.}~\bibnamefont {Lombardi}}, \bibinfo {author} {\bibfnamefont
  {F.}~\bibnamefont {Tafuri}}, \ and\ \bibinfo {author} {\bibfnamefont
  {A.}~\bibnamefont {Tagliacozzo}},\ }\bibfield  {title} {\enquote {\bibinfo
  {title} {Incipient {B}erezinskii-{K}osterlitz-{T}houless transition in
  two-dimensional coplanar {J}osephson junctions},}\ }\href {\doibase
  10.1103/PhysRevB.94.054525} {\bibfield  {journal} {\bibinfo  {journal} {Phys.
  Rev. B}\ }\textbf {\bibinfo {volume} {94}},\ \bibinfo {pages} {054525}
  (\bibinfo {year} {2016})}\BibitemShut {NoStop}%
\bibitem [{\citenamefont {S{\o}rensen}\ \emph {et~al.}(2017)\citenamefont
  {S{\o}rensen}, \citenamefont {Pedersen},\ and\ \citenamefont
  {{\"O}gren}}]{SORE17}%
  \BibitemOpen
  \bibfield  {author} {\bibinfo {author} {\bibfnamefont {M.~P.}\ \bibnamefont
  {S{\o}rensen}}, \bibinfo {author} {\bibfnamefont {N.~F.}\ \bibnamefont
  {Pedersen}}, \ and\ \bibinfo {author} {\bibfnamefont {M.}~\bibnamefont
  {{\"O}gren}},\ }\bibfield  {title} {\enquote {\bibinfo {title} {The dynamics
  of magnetic vortices in type {II} superconductors with pinning sites studied
  by the time dependent {G}inzburg--{L}andau model},}\ }\href {\doibase
  10.1016/j.physc.2016.08.001} {\bibfield  {journal} {\bibinfo  {journal}
  {Physica C}\ }\textbf {\bibinfo {volume} {533}},\ \bibinfo {pages} {40}
  (\bibinfo {year} {2017})}\BibitemShut {NoStop}%
\bibitem [{\citenamefont {Reichhardt}\ \emph {et~al.}(1998)\citenamefont
  {Reichhardt}, \citenamefont {Olson},\ and\ \citenamefont {Nori}}]{REIC98}%
  \BibitemOpen
  \bibfield  {author} {\bibinfo {author} {\bibfnamefont {C.}~\bibnamefont
  {Reichhardt}}, \bibinfo {author} {\bibfnamefont {C.~J.}\ \bibnamefont
  {Olson}}, \ and\ \bibinfo {author} {\bibfnamefont {F.}~\bibnamefont {Nori}},\
  }\bibfield  {title} {\enquote {\bibinfo {title} {Nonequilibrium dynamic
  phases and plastic flow of driven vortex lattices in superconductors with
  periodic arrays of pinning sites},}\ }\href {\doibase
  10.1103/physrevb.58.6534} {\bibfield  {journal} {\bibinfo  {journal} {Phys.
  Rev. B}\ }\textbf {\bibinfo {volume} {58}},\ \bibinfo {pages} {6534}
  (\bibinfo {year} {1998})}\BibitemShut {NoStop}%
\bibitem [{\citenamefont {Cooley}\ and\ \citenamefont
  {Grishin}(1995)}]{COOL95}%
  \BibitemOpen
  \bibfield  {author} {\bibinfo {author} {\bibfnamefont {L.~D.}\ \bibnamefont
  {Cooley}}\ and\ \bibinfo {author} {\bibfnamefont {A.~M.}\ \bibnamefont
  {Grishin}},\ }\bibfield  {title} {\enquote {\bibinfo {title} {Pinch effect in
  commensurate vortex-pin lattices},}\ }\href {\doibase
  10.1103/physrevlett.74.2788} {\bibfield  {journal} {\bibinfo  {journal}
  {Phys. Rev. Lett.}\ }\textbf {\bibinfo {volume} {74}},\ \bibinfo {pages}
  {2788} (\bibinfo {year} {1995})}\BibitemShut {NoStop}%
\bibitem [{\citenamefont {Cooley}\ \emph {et~al.}(1994)\citenamefont {Cooley},
  \citenamefont {Lee}, \citenamefont {Larbalestier},\ and\ \citenamefont
  {O'Larey}}]{COOL94}%
  \BibitemOpen
  \bibfield  {author} {\bibinfo {author} {\bibfnamefont {L.~D.}\ \bibnamefont
  {Cooley}}, \bibinfo {author} {\bibfnamefont {P.~J.}\ \bibnamefont {Lee}},
  \bibinfo {author} {\bibfnamefont {D.~C.}\ \bibnamefont {Larbalestier}}, \
  and\ \bibinfo {author} {\bibfnamefont {P.~M.}\ \bibnamefont {O'Larey}},\
  }\bibfield  {title} {\enquote {\bibinfo {title} {Periodic pin array at the
  fluxon lattice scale in a high-field superconducting wire},}\ }\href
  {\doibase 10.1063/1.111940} {\bibfield  {journal} {\bibinfo  {journal} {Appl.
  Phys. Lett.}\ }\textbf {\bibinfo {volume} {64}},\ \bibinfo {pages}
  {1298} (\bibinfo {year} {1994})}\BibitemShut {NoStop}%
\bibitem [{\citenamefont {Reichhardt}\ \emph {et~al.}(1997)\citenamefont
  {Reichhardt}, \citenamefont {Groth}, \citenamefont {Olson}, \citenamefont
  {Field},\ and\ \citenamefont {Nori}}]{REIC97}%
  \BibitemOpen
  \bibfield  {author} {\bibinfo {author} {\bibfnamefont {C.}~\bibnamefont
  {Reichhardt}}, \bibinfo {author} {\bibfnamefont {J.}~\bibnamefont {Groth}},
  \bibinfo {author} {\bibfnamefont {C.~J.}\ \bibnamefont {Olson}}, \bibinfo
  {author} {\bibfnamefont {S.~B.}\ \bibnamefont {Field}}, \ and\ \bibinfo
  {author} {\bibfnamefont {F.}~\bibnamefont {Nori}},\ }\bibfield  {title}
  {\enquote {\bibinfo {title} {Spatiotemporal dynamics and plastic flow of
  vortices in superconductors with periodic arrays of pinning sites},}\ }\href
  {\doibase 10.1103/physrevb.54.16108} {\bibfield  {journal} {\bibinfo
  {journal} {Phys. Rev. B}\ }\textbf {\bibinfo {volume} {54}},\ \bibinfo
  {pages} {16108} (\bibinfo {year} {1997})}\BibitemShut {NoStop}%
\bibitem [{\citenamefont {Eder-Kapl}\ \emph {et~al.}(2012)\citenamefont
  {Eder-Kapl}, \citenamefont {Steiger-Thirsfeld}, \citenamefont {Wellenzohn},
  \citenamefont {Koeck}, \citenamefont {H.}, \citenamefont {Loeschner},\ and\
  \citenamefont {Platzgummer}}]{EDER12}%
  \BibitemOpen
  \bibfield  {author} {\bibinfo {author} {\bibfnamefont {S.}~\bibnamefont
  {Eder-Kapl}}, \bibinfo {author} {\bibfnamefont {A.}~\bibnamefont
  {Steiger-Thirsfeld}}, \bibinfo {author} {\bibfnamefont {M.}~\bibnamefont
  {Wellenzohn}}, \bibinfo {author} {\bibfnamefont {A.}~\bibnamefont {Koeck}},
  \bibinfo {author} {\bibfnamefont {Rainer}\ \bibnamefont {H.}}, \bibinfo
  {author} {\bibfnamefont {H.}~\bibnamefont {Loeschner}}, \ and\ \bibinfo
  {author} {\bibfnamefont {E.}~\bibnamefont {Platzgummer}},\ }\bibfield
  {title} {\enquote {\bibinfo {title} {Ion multi-beam direct sputtering of {S}i
  imprint stamps and simulation of resulting structures},}\ }\href {\doibase
  10.1088/0960-1317/22/5/055008} {\bibfield  {journal} {\bibinfo  {journal} {J.
  Micromech. Microeng.}\ }\textbf {\bibinfo {volume} {22}},\ \bibinfo {pages}
  {055008} (\bibinfo {year} {2012})}\BibitemShut {NoStop}%
\end{thebibliography}
\end{document}